\documentclass[useAMS,usegraphicx,usenatbib]{mn2e} 
\usepackage{amsmath,graphicx}
\usepackage{ulem}
\usepackage{color}
\usepackage{graphicx}
\usepackage{times} 
\usepackage{amssymb}
\usepackage{amsmath}
\usepackage{lscape}
\usepackage{url}
\usepackage{multirow}
\usepackage{longtable}
\DeclareGraphicsExtensions{.pdf,.png,.jpg,.eps}
\newif\ifAMStwofonts
\AMStwofontstrue
\newcommand{\kpc}{\rm\thinspace kpc}

\newcommand{\km}{\rm\thinspace km}

\newcommand{\cm}{\rm\thinspace cm}

\newcommand{\yr}{\rm\thinspace yr}

\newcommand{\Myr}{\rm\thinspace Myr}
\newcommand{\s}{\rm\thinspace s}

\newcommand{\Msun}{\hbox{$\rm\thinspace M_{\odot}$}}

\newcommand{\Msunpyr}{\hbox{$\Msun\yr^{-1}$}}
\newcommand{\kmps}{\hbox{$\km\s^{-1}\,$}}

\newcommand{\pcmcu}{\hbox{$\cm^{-3}\,$}}

\def\h0{\mbox{{\rm H}$^0$}}
\def\he0{\mbox{{\rm He}$^0$}}
\begin{document}

\title[Filamentary Star Formation in NGC 1275] {Filamentary Star Formation in NGC 1275} \author[]  {\parbox[]{6.in} %\author[R. E. A. Canning et al.]
{R.E.A.~Canning,$^{1,2}$\thanks{E-mail: rcanning@stanford.edu} J.E. Ryon,$^{3}$ J.S. Gallagher III,$^{3}$ R. Kotulla,$^{4}$ R.W. O'Connell,$^{5}$ A.C. Fabian,$^{6}$, R.M. Johnstone,$^{6}$ C.J. Conselice,$^{7}$ A. Hicks,$^{8}$ D. Rosario,$^{9}$ and R.F.G. Wyse$^{10}$} \\ \\
\footnotesize 
$^{1}$Kavli Institute for Particle Astrophysics and Cosmology (KIPAC), Stanford University, 452 Lomita Mall, Stanford, CA 94305-4085, USA\\
$^{2}$Department of Physics, Stanford University, 452 Lomita Mall, Stanford, CA 94305-4085, USA\\
$^{3}$Department of Astronomy, University of Wisconsin-Madison, 475 N. Charter St., Madison, WI, 53706 \\
$^{4}$Department of Physics, University of Wisconsin-Milwaukee, 1900 E. Kenwood Blvd., Milwaukee, WI, 53211\\
$^{5}$Department of Astronomy, University of Virginia, P.O. Box 400235, Charlottesville, VI, 22904\\
$^{6}$Institute of Astronomy, University of Cambridge, Madingley Road, Cambridge CB3 0HA\\
$^{7}$School of Physics \& Astronomy, University of Nottingham, University Park, Nottingham NG7 2RD\\
$^{8}$Department of Physics \& Astronomy, Michigan State University, East Lansing, MI 48824-2320, USA\\
$^{9}$Max Planck Institute for Extraterrestrial Physics, Postfach 1312, 85741 Garching, Germany\\
$^{10}$Physics and Astronomy Department, Johns Hopkins University, 3400 North Charles Street, Baltimore, MD 21218, USA\\
}

\maketitle

\label{firstpage}

\begin{abstract} 
We examine the star formation in the outer halo of NGC~1275, the central galaxy in the Perseus cluster (Abell 426), using far ultraviolet and optical images obtained with the {\it Hubble Space Telescope}. We have identified a population of very young, compact star clusters with typical ages of a few~Myr. The star clusters are organized on multiple-kiloparsec scales. Many of these star clusters are associated with ``streaks'' of young stars, the combination of which has a cometary appearance. We perform photometry on the star clusters and diffuse stellar streaks, and fit their spectral energy distributions to obtain ages and masses. These young stellar populations appear to be normal in terms of their masses, luminosities and cluster formation efficiency; $<$10\% of the young stellar mass is located in star clusters. Our data suggest star formation is associated with the evolution of some of the giant gas filaments in NGC~1275 that become gravitationally unstable on reaching and possibly stalling in the outer galaxy. The stellar streaks then could represent stars moving on ballistic orbits in the potential well of the galaxy cluster. We propose a model where star-forming filaments, switched on $\sim$50~Myr ago and are currently feeding the growth of the NGC~1275 stellar halo at a rate of $\approx$2-3~M$_{\odot}$~yr$^{-1}$. This type of process may also build stellar halos and form isolated star clusters in the outskirts of youthful galaxies.
\end{abstract}


\section{Introduction}
\label{intro}

Most star formation in the local universe occurs in galaxy disks. Dynamical instabilities in disks aid in the formation of star-forming gas clouds \citep[e.g.,][]{toomre1977,kennicutt1998,leroy2008}.  Low level star formation is also observed in more diffuse gas structures, including extraplanar colliding gas shells \citep{tullmann2003}, tidal debris \citep{demello2008,werk2008}, and in gas on the boundaries of radio jets and lobes \citep[e.g.,][]{graham1998,mould2000,odea2004}.  The most spectacular examples of star formation outside of galactic disks, however, occur in the vicinities of some brightest cluster galaxies (BCGs) in rich X-ray `cool core' galaxy clusters \citep[e.g.][]{allen1995,mcnamara2006, rafferty2008,odea2008,quillen2008,odea2010,hicks2010,mcdonald2011,oonk2011, tremblay2012, mcdonald2012}.  These objects can support galactic-scale star formation in extended gas structures which are products of interactions between the BCG and its surroundings.

In this paper, we present an ultraviolet imaging study of the spectacular outer star-forming filaments associated with NGC~1275, the BCG in the Perseus galaxy cluster, based on data obtained with the {\it Hubble Space Telescope} (\textit{HST}). Despite being initially classified as an early-type galaxy \citep{hubble1931}, several structural oddities of NGC~1275 were soon discovered. One mysterious feature is an extensive ``web" of gaseous, emission line filaments, strongly emitting in H$\alpha$ \citep{minkowski1955, lynds1970}.  Observations of H$_{2}$ and CO have shown a substantial mass of molecular gas exists in the core of NGC 1275 \citep{bridges1998,donahue2000,edge2002} and is also entrained in the filaments \citep{hatch2005, salome2006,johnstone2007b,salome2011,lim2012}, and recent far infrared (FIR) observations with the {\it Herschel} space telescope are allowing us to probe the nature of this cold gas reservoir \citep{mittal2012}. The most promising mechanism for heating these multiphase filaments is supra-thermal particle heating from the surrounding hot X-ray ICM \citep{ferland2009}.

Perhaps the most peculiar feature of NGC~1275 is evidence for recent massive star formation in and around the galaxy's main body. The presence of this star formation has been known for some time \citep[see e.g.][and references therein]{sandage1971, vandenbergh1977}. \cite{adams1977} noted ultraviolet (UV) emission beyond the galaxy's main body to the northwest and southeast. He suggested it could be due to the presence of hot stars, a point also discussed by \cite{smith1992} on the basis of far ultraviolet (FUV) imaging. Based on stellar population synthesis modeling and the detection of strong Balmer emission in an integrated spectrum of the central 25\arcsec\ of NGC~1275, \cite{wirth1983} concluded that a large population of B stars must be present. A localized population of young stars was directly detected spectroscopically $\sim$19\arcsec\ from the galaxy's center by \cite{shields1990}, and since then, multiple young clusters have been found in the galaxy's outskirts (e.g., \citealt{shields1990, holtzman1992}). \cite{mcnamara1996} conducted a multiwavelength study of the structure of NGC~1275, and suggested the presence of young stars in the outskirts, which were later found by \cite{conselice2001} to have very blue colors, consistent with young stellar populations.

\cite{canning2010a} (hereafter Paper I) used optical HST imaging to demonstrate that the outer UV-bright structures contain star clusters with ages $<$100 Myr, in contrast to the high-luminosity clusters in the galaxy's central regions with ages $>$100 Myr. The authors infer, from the optical colors, a SFR of $\sim$20~\Msunpyr in these outer regions and conclude that the outer stellar features must have a different origin to the blue cluster population studied by \cite{carlson1998} in the central regions of NGC 1275. However, without higher resolution UV imaging, more accurate ages of these star clusters could not determined. Paper I also discusses the presence of clumpy stellar ``streaks", cometary features with a star cluster  at the ``head," followed by a ``tail" of more diffuse stellar light, within the larger UV-bright structures of \cite{adams1977} (see section \ref{prop_diffuse_sources}). These streaks may be due to the dissolution of the star clusters as they are released from their natal gas and fall toward the center of NGC~1275. More accurate ages of the star clusters and stellar streaks are needed to understand their origin. 

Our investigation utilizes results of our FUV and optical \textit{HST} ACS observations of the outer UV-bright filaments of young stars in NGC~1275. We combine Solar Blind Channel (SBC) FUV data with Wide Field Camera (WFC) optical data to obtain accurate ages and masses of the young star cluster candidates and associated diffuse light. Section~\ref{observations} describes our \textit{HST} observations, data reduction, source selection, and photometry. In Section~\ref{galev}, we describe the method for fitting the SEDs of our sources. Section~\ref{prop_star_clusters} addresses the colors, masses, and ages of the compact star clusters.  In Section~\ref{prop_diffuse_sources}, we discuss the properties of the stellar streaks. Section~\ref{discussion} addresses how the star formation studied here fits in the context of the large-scale emission-line filaments. Section~\ref{conclusions} contains our conclusions.

Throughout this paper we assume the standard $\Lambda$CDM cosmology where H$_{0}=71$~km~s$^{-1}$, $\Omega_{m}=0.3$ and $\Omega_{\lambda}=0.7$. For this cosmology and at the redshift of NGC~1275 ($z=0.0176$), an angular size of 10\arcsec\ corresponds to a distance of 3.5 kpc.

\section{Observations \& Data Reduction}
\label{observations}

\begin{figure*}
 \includegraphics[width=\textwidth]{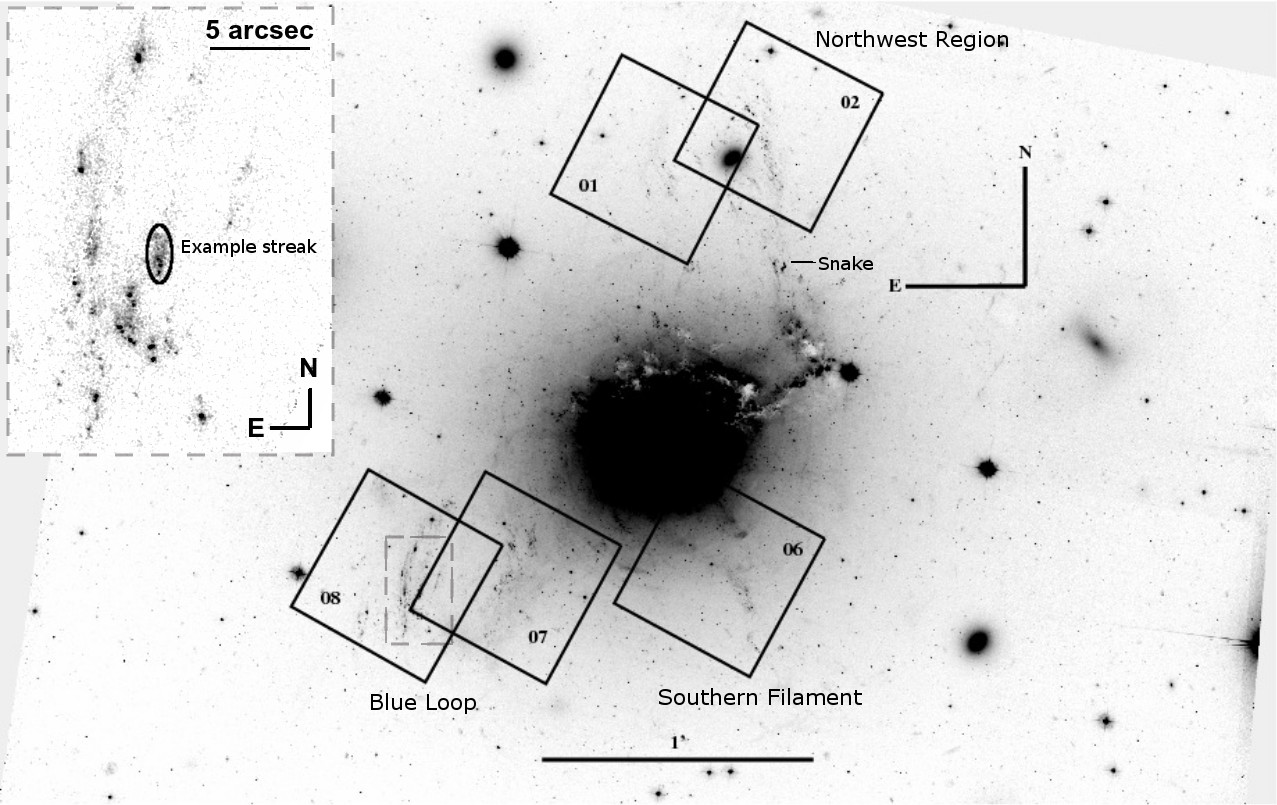}
 \caption{F435W optical image overlaid with the five SBC pointings analysed in this paper. 01 and 02 cover the Northwest Region, 06 covers the Southern Filament, and 07 and 08 cover the Blue Loop. The inset shows a close up of the eastern arm of the Blue Loop (see pointing 08) illustrating the presence of diffuse stellar ``streaks'' and compact UV sources; one such streak is labeled. The SBC field-of-view covers 34.6''$\times$30.8'' while the larger ACS WFC field-of-view is 202''$\times$202''. The center of NGC 1275 is at coordinates 49.9507$^{\circ}$, 41.5118$^{\circ}$.\label{sbc_pointings}}
\end{figure*}

Optical observations of NGC~1275 were obtained with the \textit{HST} ACS/WFC on 2006 August 5 in the F435W, F550M, and F625W filters (first published in \citealt{fabian2008}), The ACS/WFC field of view is 202''$\times$202'' with pixel scale 0.05''/pixel. 2 F435W pointings and 3 F550M and F6255W pointings were required, taken in a mosaic across NGC 1275, and FUV observations were obtained with the ACS/SBC on 2008 March 16 to 21 in the F140LP filter. Table~\ref{obs_summary} summarizes these observations. The field of view of ACS/SBC (FoV: 34.6''$\times$30.8'', pixel scale: 0.034''$\times$0.030''/pixel) is smaller than the ACS/WFC and multiple pointings with the SBC were required to image the sites of outer star formation. Figure~\ref{sbc_pointings} shows these SBC pointing locations overlaid on the WFC F435W optical image. Note that the SBC pointings have been labeled 01, 02, 06, 07, and 08. Pointings 03 and 04 suffer from confusion with the HVS and so we choose not to analyse these images here. 01 and 02 cover the Northwest Region, 06 covers the Southern Filament, and 07 and 08 cover the Blue Loop.

NGC~1275 consists of two distinct velocity components, the high velocity system (HVS) at $v=8200$~\kmps\ and the low velocity system (LVS) at $v=5200$~\kmps\ \citep{minkowski1955}. The HVS is associated with a dusty foreground object, possibly a late-type galaxy falling in to the cluster, while the early-type galaxy and emission-line filaments constitute the LVS \citep{rubin1977,conselice2001}. X-ray absorption by the HVS requires that the distance between the HVS and LVS is greater than 100 kpc \citep{sanders2007}. The star formation in the Northwest Region (pointings 01 and 02) could be associated with the HVS, but we view this as unlikely due to the morphological similarity of the stellar streaks with the emission-line filaments. In addition, spectroscopic studies have confirmed that a complex of bright star clusters to the south of the Northwest Region, nicknamed the `Snake' (see Figure \ref{sbc_pointings}), lies at the redshift of the LVS \citep{hatch2006}. For the remainder of this paper we shall make the assumption that the star formation identified in the regions specified above are associated with the LVS.

\begin{figure*}
  \centering
  \includegraphics[width=0.9\textwidth]{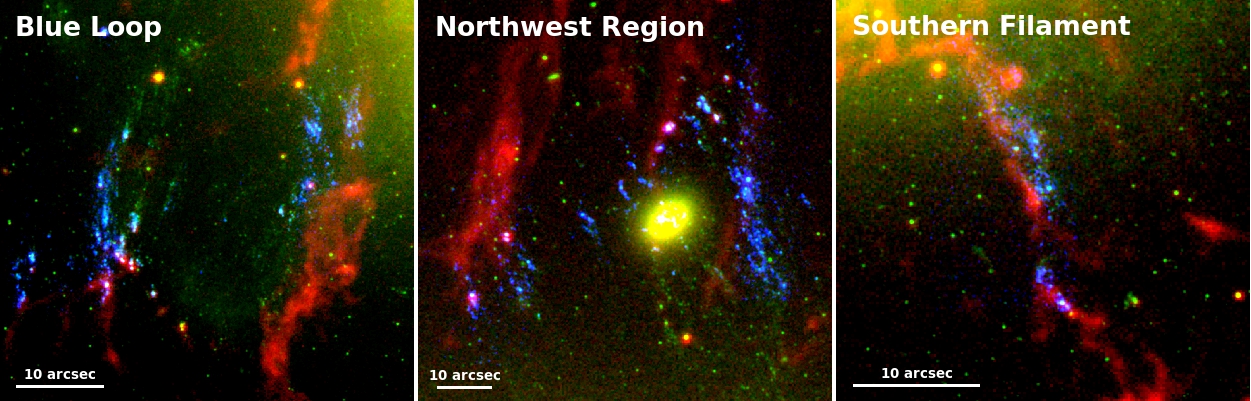}
  \caption{Three color H$\alpha$, F435W and F140LP images of the three star forming regions in the outskirts of the NGC 1275 low-velocity system; (left) Blue Loop (49.9671$^{\circ}$, 41.5040$^{\circ}$), (middle) Northwest Region (49.9458$^{\circ}$,41.5295$^{\circ}$), and (right) Southern Filament (49.4937$^{\circ}$, 41.5019$^{\circ}$). The H$\alpha$ is from narrow-band imaging using the WIYN telescope, The bright galaxy in the Northern Region is a cluster member 2MASXi J0319467+413145. }\label{Ha_contours}
\end{figure*}

The WFC F625W filter includes H$\alpha$, [N {\sc ii}] and [S {\sc ii}] emission from both the HVS and LVS and the WFC F435W filter includes [O {\sc ii}] emission from both the HVS and LVS. The SBC F140LP filter transmits both the two photon continuum, which is expected to be associated with the H$\alpha$ filaments \citep{johnstone2012}, and C {\sc iv}~1549~\AA\ line emission, a strong cooling line diagnostic of 10$^{5}$ K gas which has been seen in a similar filamentary system surrounding M87 in the Virgo cluster \citep{sparks2009, sparks2012}. As noted in Paper I, the young star clusters are typically offset from the brightest filamentary emission (typical offsets are $\sim$ a few arcsec so 0.6-1 kpc; see Figure \ref{Ha_contours}); the distribution of the UV emission follows that of the broad band optical filters not the narrow-band H$\alpha$ imaging.  Using WIYN narrow-band H$\alpha$ imaging and scaling from the ratios of C{\sc iv}, He{\sc ii} \citep{sparks2009, sparks2012} and $F_{1500\mathrm{\AA}}$ \citep{johnstone2012} to H$\alpha$ on the tails of the streaks we estimate emission line contamination in the F140LP filter to be at most 2 thousandths of the total flux. However, in order to minimise the contamination by emission lines, not associated with the clusters, we take background regions local to the sources for our photometry. Our approach therefore assumes that any background due to ionized gas and stars is uniform over small regions of the outer galaxy.

The standard procedure was used to bias-subtract and flat-field the data frames. These were then drizzled together as described in the HST ACS data handbook\footnote{http://www.stsci.edu/hst/acs/documents/handbooks/currentDHB}. The SBC suffers from a red leak. Accurate removal of the contaminating light from our data requires additional exposures of NGC~1275 using a redder SBC filter. However, these data currently do not exist. Fortunately, A-type stars and earlier have very minimal contamination at these wavelengths (see HST/ACS Data Handbook$^{1}$). Additionally, our analysis focuses on the outer regions ($\sim$20 kpc from the nucleus) of NGC~1275 where the underlying red galaxy is not very luminous. We therefore conclude that the red leak contribution to the emission measured in the vicinity of the UV-bright sources in these regions is negligible.

Both very diffuse and compact FUV emission is observed in these three regions. In this paper we ignore the most diffuse component of the UV emission and concentrate on the properties of the cometary stellar ``streaks''. In doing so we are measuring a lower limit for the mass of young stars in these regions.
The insert in Figure~\ref{sbc_pointings} illustrates the morphology of a stellar ``streak''. The UV emission has a cometary appearance, with a compact star cluster at the ``head'' and more diffuse emission following in a ``tail''.  In all regions the ``heads'' are orientated away from the central galaxy. We used different techniques to perform photometry on these components. Some of the streaks are not obviously associated with a star cluster, but we included them in this analysis nevertheless.

We used IRAF\footnote{IRAF is distributed by the National Optical Astronomy Observatories, which are operated by the Association of Universities for Research in Astronomy, Inc., under cooperative agreement with the National Science Foundation.} DAOFIND to perform source selection on the compact sources in the SBC images. We used a threshold of 13 times the standard deviation of the background, as determined in regions of each image without bright sources. We constrained the FWHM to 6 pixels, and the sharpness and roundness parameters to 0.2 to 0.8 and -1 to 1, respectively. A few obvious sources missed by DAOFIND were added in by hand through visual inspection of the images. Spurious sources, including chip edge effects, and real sources detected in overlapping regions of the SBC pointings were removed from the catalogue. The catalogue was then matched to corresponding sources in the optical; no source selection was performed on the WFC images. The above procedure strongly biases us towards selecting very young star clusters, since young, massive stars are the main stellar source of FUV flux. We justify this by noting that we are primarily interested in sites of recent or ongoing star formation. We will call these compact sources star clusters for the rest of this paper, although we could not determine if they are bound or unbound.

\begin{table}
 \centering
  \caption{Summary of \textit{HST} Observations \label{obs_summary}}
  \begin{tabular}{ccccc}
  \hline
Camera & Filter & Date & Program &  Exp. Time \\
             &          &         &      ID     &    (s)  \\
 \hline
 ACS/SBC & F140LP & 16-21 March 2008 & GO 11207 & 2552\\
ACS/WFC & F435W & 05 August 2006 & GO 10546 & 9834\\
ACS/WFC & F550M & 05 August 2006 & GO 10546 & 12,132 \\
ACS/WFC & F625W & 05 August 2006 & GO 10546 & 12,405 \\
\hline
\end{tabular}
\end{table}

We performed aperture photometry on our catalogue with DAOPHOT. The STMAG zeropoints were determined by the procedure given in the \textit{HST} ACS Data Handbook. We chose photometric apertures of radius 0.1'', which corresponds to a physical radius of $\sim$35~pc at the distance of NGC~1275. The small apertures avoid overlap of sources in crowded regions. Growth curves of bright, isolated sources were constructed to find optimal sky annuli. Average aperture corrections for the WFC data were determined in each region from isolated clusters with smooth growth curves. For the SBC data, empirical aperture corrections were deemed unreliable due to strongly varying growth curves. We therefore derived aperture corrections from the encircled energy fractions of \cite{dieball2007}. Table~\ref{aperture_phot} contains our photometry parameters and aperture corrections. The magnitudes and magnitude errors, with aperture corrections applied, of the star clusters in our catalog are listed in Table~\ref{photometry}. The aperture corrections dominate the magnitude uncertainties. Note that these magnitudes have been corrected for the foreground extinction listed in Table~\ref{aperture_phot} using the reddening law of \cite{cardelli1989}. 

\begin{table*}
\centering
\caption{Aperture Photometry Parameters and Aperture Corrections. \label{aperture_phot} }
\begin{tabular}{ccccccccc}
\hline
Camera \& & Zeropoint & Aperture & Annular & Annular & Foreground & & Aperture &  \\
  Filter & (STMAG) & (pix) & Radius & Width & Extinction &  & Correction$^a$ (mag)  &  \\
 & & &  (pix) & (pix) & (mag) & A & B & C \\
\hline
SBC F140LP & 20.3164 & 4 & 7 & 5 & 1.35 & $0.67$ &  $0.67$ & $0.67$ \\
WFC F435W & 25.1572 & 2 & 15 & 10 & 0.67 & $0.51\pm0.05$ & $0.54\pm0.07$ & $0.6\pm0.1$ \\
WFC F550M & 24.9335 & 2 & 15 & 10 & 0.50 & $0.54\pm0.04$ & $0.5\pm0.1$ & $0.57\pm0.05$ \\
WFC F625W & 26.2062 & 2 & 15 & 10 & 0.43 & $0.68\pm0.07$ & $0.71\pm0.07$ & $0.58\pm0.07$ \\
\hline
\multicolumn{8}{l}{
  \begin{minipage}{14cm}
  \textit{Note:} A, B, and C refer to the Northwest Region, Southern Filament, and Blue Loop, respectively. \\
  $^a$ Aperture corrections calculated from the average difference in flux between apertures of 0.1\arcsec\ radius and 0.6\arcsec\ radius.
  \end{minipage}}
\end{tabular}
\end{table*}

\begin{table*}
\centering
\caption{Star Cluster Photometry and Derived Properties \label{photometry}  }
\begin{tabular}{ccccccccccc}
\hline
RA & Dec & F140LP$_{0}$ & F435W$_{0}$ & F550M$_{0}$ & F625W$_{0}$ & Age  & 1~$\sigma$~range & Mass & 1$\sigma$~range \\
J2000 & J2000 & (mag) & (mag) & (mag) & (mag) & (Myr) & (Myr) & ($10^{3}$~M$_{\odot}$) & ($10^{3}$~M$_{\odot}$)\\
\hline
 \multicolumn{10}{l}{Northwest Region} \\
\hline
3:19:49.1 & +41:31:32 & 20.99$\pm$ 0.07 & 24.17$\pm$ 0.05 & 24.85$\pm$ 0.05 & 24.89$\pm$ 0.07 &   4.0 & [2.3, 4.1] & 12.1 & [7.25, 12.4] \\
3:19:48.4 & +41:31:37 & 20.50$\pm$ 0.06 & 23.60$\pm$ 0.05 & 24.47$\pm$ 0.04 & 24.01$\pm$ 0.07 &   3.9 & [1.9, 4.1] & 16.4 & [10.8, 17.5] \\
3:19:47.9 & +41:31:41 & 21.78$\pm$ 0.11 & 24.43$\pm$ 0.05 & 25.09$\pm$ 0.05 & 25.19$\pm$ 0.07 &  11.2 & [2.7, 12.6] & 29.4 & [5.71, 33.7] \\
 3:19:46.8 & +41:31:51 & 21.40$\pm$ 0.09 & 24.48$\pm$ 0.05 & 25.33$\pm$ 0.05 & 25.31$\pm$ 0.07 &   4.0 & [2.2, 4.0] & 8.33 & [4.79, 8.33] \\
 3:19:46.2 & +41:32:01 & 21.37$\pm$ 0.09 & 24.47$\pm$ 0.05 & 25.43$\pm$ 0.05 & 25.58$\pm$ 0.07 &   4.0 & [2.1, 4.0] & 8.46 & [4.79, 8.46] \\
 3:19:46.5 & +41:31:56 & 21.02$\pm$ 0.08 & 24.05$\pm$ 0.05 & 25.00$\pm$ 0.05 & 25.25$\pm$ 0.07 &   4.0 & [2.2, 4.0] & 11.8 & [6.84, 11.8] \\
 3:19:46.2 & +41:31:58 & 21.15$\pm$ 0.08 & 24.11$\pm$ 0.05 & 25.12$\pm$ 0.05 & 24.96$\pm$ 0.07 &   2.4 & [2.2, 4.0] & 6.52 & [5.97, 10.6] \\
 3:19:46.3 & +41:31:54 & 20.58$\pm$ 0.06 & 22.92$\pm$ 0.05 & 23.62$\pm$ 0.04 & 23.66$\pm$ 0.07 &   4.2 & [2.5, 17.6] & 38.5 & [24.4, 193.0] \\
 3:19:46.0 & +41:31:48 & 20.26$\pm$ 0.05 & 22.98$\pm$ 0.05 & 23.79$\pm$ 0.04 & 24.01$\pm$ 0.07 &   4.1 & [2.5, 8.9] & 28.3 & [16.3, 97.5] \\
 3:19:46.1 & +41:31:48 & 21.32$\pm$ 0.09 & 24.26$\pm$ 0.06 & 25.12$\pm$ 0.05 & 25.34$\pm$ 0.07 &   4.0 & [2.3, 4.1] & 9.07 & [5.58, 9.27] \\
 3:19:46.4 & +41:31:55 & 21.10$\pm$ 0.08 & 23.68$\pm$ 0.05 & 24.37$\pm$ 0.04 & 24.58$\pm$ 0.07 &   4.3 & [3.1, 12.0] & 15.5 & [10.6, 60.0] \\
 3:19:46.2 & +41:31:52 & 21.32$\pm$ 0.09 & 24.32$\pm$ 0.05 & 25.17$\pm$ 0.05 & 25.48$\pm$ 0.07 &   4.0 & [2.4, 4.1] & 8.99 & [5.53, 9.17] \\
 3:19:46.4 & +41:31:56 & 21.29$\pm$ 0.09 & 24.14$\pm$ 0.05 & 25.02$\pm$ 0.05 & 25.23$\pm$ 0.07 &   4.0 & [2.2, 8.0] & 10.3 & [6.05, 31.8] \\
 3:19:46.7 & +41:31:53 & 21.26$\pm$ 0.09 & 23.86$\pm$ 0.05 & 24.74$\pm$ 0.05 & 24.16$\pm$ 0.07 &   3.6 & [0.2, 8.1] & 15.7 & [9.31, 29.4]\\
 \hline
 \multicolumn{10}{l}{Southern Filament} \\
 \hline
 3:19:46.2 & +41:29:57 & 20.90$\pm$0.07 & 23.40$\pm$0.07 & 24.27$\pm$0.10 & 24.20$\pm$0.07 & 4.7 & [2.3, 12.6] & 17.7 & [12.5, 77.8] \\
 \hline
 \multicolumn{10}{l}{Blue Loop}  \\
 \hline
  3:19:51.0 & +41:30:19 & 21.29$\pm$ 0.09 & 24.01$\pm$ 0.10 & 25.09$\pm$ 0.07 & 25.31$\pm$ 0.08 &   4.0 & [2.3, 4.2] & 9.73 & [5.71, 9.73] \\
 3:19:51.1 & +41:30:12 & 20.50$\pm$ 0.06 & 23.33$\pm$ 0.10 & 24.35$\pm$ 0.05 & 23.70$\pm$ 0.07 &   2.0 & [1.4, 4.1] & 16.2 & [12.8, 22.4] \\
 3:19:50.7 & +41:30:19 & 21.48$\pm$ 0.10 & 23.66$\pm$ 0.10 & 24.45$\pm$ 0.06 & 24.77$\pm$ 0.08 &  19.9 & [3.0, 21.5] & 64.4 & [11.0, 76.7] \\
 3:19:51.4 & +41:30:09 & 21.89$\pm$ 0.12 & 24.59$\pm$ 0.10 & 25.44$\pm$ 0.08 & 25.84$\pm$ 0.09 &   4.1 & [2.5, 8.8] & 5.64 & [3.59, 21.1] \\
 3:19:53.9 & +41:30:03 & 21.11$\pm$ 0.08 & 23.77$\pm$ 0.10 & 24.71$\pm$ 0.05 & 24.32$\pm$ 0.07 &   4.0 & [1.6, 11.6] & 16.8 & [9.07, 56.5] \\
 3:19:54.0 & +41:30:00 & 21.72$\pm$ 0.11 & 24.49$\pm$ 0.10 & 25.53$\pm$ 0.07 & 24.99$\pm$ 0.07 &   3.6 & [1.1, 4.4] & 6.46 & [4.03, 7.87] \\
 3:19:53.0 & +41:30:18 & 20.41$\pm$ 0.06 & 22.95$\pm$ 0.10 & 23.90$\pm$ 0.05 & 24.13$\pm$ 0.07 &   3.2 & [2.5, 8.9] & 17.2 & [14.6, 86.0] \\
 3:19:53.2 & +41:30:12 & 20.25$\pm$ 0.05 & 22.84$\pm$ 0.10 & 23.78$\pm$ 0.05 & 24.05$\pm$ 0.07 &   3.2 & [2.5, 8.6] & 19.8 & [16.7, 94.5] \\
 3:19:53.2 & +41:30:07 & 20.84$\pm$ 0.07 & 23.60$\pm$ 0.10 & 24.57$\pm$ 0.06 & 24.58$\pm$ 0.07 &   4.0 & [2.1, 8.1] & 17.8 & [9.34, 49.9] \\
 3:19:53.0 & +41:30:06 & 20.53$\pm$ 0.06 & 23.33$\pm$ 0.10 & 24.32$\pm$ 0.05 & 24.32$\pm$ 0.07 &   4.0 & [2.2, 4.3] & 21.9 & [12.3, 21.9] \\
 3:19:53.0 & +41:30:05 & 20.72$\pm$ 0.07 & 23.52$\pm$ 0.10 & 24.54$\pm$ 0.06 & 24.10$\pm$ 0.07 &   3.9 & [1.6, 4.6] & 17.6 & [9.74, 18.1] \\
 3:19:52.9 & +41:30:07 & 21.19$\pm$ 0.09 & 23.90$\pm$ 0.10 & 24.87$\pm$ 0.06 & 24.50$\pm$ 0.07 &   4.0 & [1.6, 11.4] & 13.7 & [7.40, 50.9] \\
 3:19:53.1 & +41:30:01 & 20.76$\pm$ 0.07 & 22.65$\pm$ 0.10 & 23.33$\pm$ 0.05 & 23.52$\pm$ 0.07 &   6.7 & [3.1, 34.8] & 56.9 & [23.0, 293.0] \\
 3:19:53.0 & +41:30:04 & 20.67$\pm$ 0.07 & 23.31$\pm$ 0.10 & 24.37$\pm$ 0.05 & 23.78$\pm$ 0.07 &   2.1 & [1.0, 4.3] & 18.1 & [13.3, 23.3] \\
 3:19:53.0 & +41:30:04 & 21.20$\pm$ 0.09 & 23.86$\pm$ 0.10 & 24.94$\pm$ 0.06 & 24.10$\pm$ 0.07 &   2.8 & [0.7, 3.9] & 9.94 & [5.67, 11.5] \\
 3:19:52.9 & +41:30:03 & 20.34$\pm$ 0.06 & 23.08$\pm$ 0.10 & 24.13$\pm$ 0.05 & 23.65$\pm$ 0.07 &   3.8 & [1.5, 4.7] & 25.6 & [14.7, 27.8] \\
 3:19:52.7 & +41:30:00 & 20.07$\pm$ 0.05 & 22.91$\pm$ 0.10 & 23.87$\pm$ 0.05 & 24.02$\pm$ 0.07 &   4.0 &  [2.3, 4.1] & 30.1 & [18.1, 30.1] \\
 3:19:53.9 & +41:30:04 & 21.36$\pm$ 0.09 & 23.32$\pm$ 0.10 & 23.85$\pm$ 0.05 & 24.08$\pm$ 0.07 &  10.8 & [3.4, 38.6] & 54.4 & [18.5, 171.0] \\
 3:19:52.9 & +41:30:04 & 20.49$\pm$ 0.06 & 23.20$\pm$ 0.10 & 24.17$\pm$ 0.05 & 24.03$\pm$ 0.07 &   3.4 & [2.0, 8.5] & 17.7 & [13.4, 74.1] \\
 3:19:53.2 & +41:30:30 & 21.19$\pm$ 0.09 & 23.75$\pm$ 0.10 & 24.85$\pm$ 0.06 & 24.23$\pm$ 0.07 &   2.6 & [0.9, 4.3] & 11.0 & [6.67, 14.9] \\
\hline
\end{tabular}
\end{table*}

\begin{figure*}
\centering
 \includegraphics[width=\textwidth]{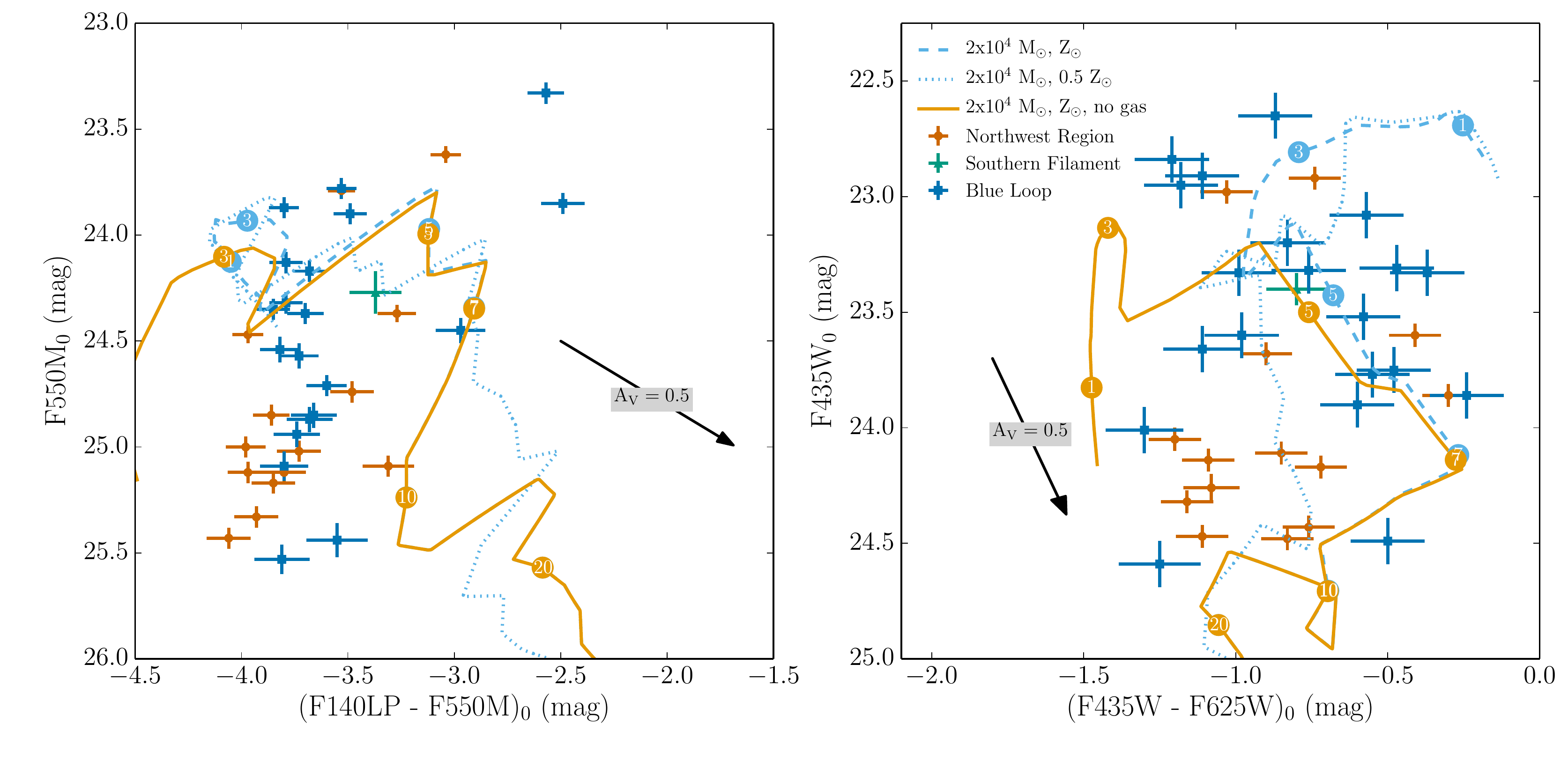}%{cmd_nogas.pdf}
 \caption{(Left) FUV-optical and (right) optical-only cluster CMDs with foreground reddening corrections applied to the measured magnitudes. The Northwest Region, Southern Filament, and Blue Loop sources are denoted with orange diamonds, green triangles, and blue squares, respectively. Blue dashed and dotted lines are evolutionary tracks corresponding to solar and half solar metallicity SEDs respectively, with mass $2\times10^4$ $\Msun$ (lowering the mass of the cluster results in the vertical movement of the tracks to dimmer magnitudes) and the orange solid line indicates the tracks for a solar metallicity model without line emission. The numbered points on each track represent the location of an SSP of age 1, 3, 5, 7, 10, and 20~\Myr. When emission lines are included initially the light is dominated by nebular continuum and H$\alpha$ emission making the F435W-F625W color redder. The turn over at 7 Myr is due to the appearance of super-giant stars, in this case the blue super-giants dominate over the red super-giant stars.
 \label{cluster_cmd}}
\end{figure*}

To estimate the contribution of the ``tails'' to the total FUV streak emission, we used SExtractor\footnote{http://www.astromatic.net/software/sextractor} \citep{bertin1996} to identify stellar streaks 5 sigma above the background in the SBC images. We determined the background by creating histograms of the pixel values in neighboring areas of the image, offset in angle, but at the same galactocentric radius as the streak under study. We tested the background on a per pixel basis and also over contiguous regions of the same size as each of our detected ellipses. We were careful to avoid areas that obviously contained star clusters but we did not excise regions where there is no star formation but which include emission-line filaments. The filaments are thin thread-like structures which add negligibly to the background. Most star formation is offset from the emission-line filaments and local backgrounds are taken for the photometry. The variance in the background was determined from the one sigma deviation of the background distribution. We then performed photometry using the apertures identified by SExtractor.

We also removed the aperture-corrected flux of the compact star cluster associated with each streak, if one was present. The apertures were sufficiently large such that the emission from each streak was encapsulated within the aperture, so we did not apply aperture corrections. Using circular apertures of radius 0.15'' placed by eye along the streaks, we also investigate the possibility of an age gradient. This is discussed further in Section~\ref{prop_diffuse_sources}. The background was determined in the same fashion as for the elliptical apertures encompassing the diffuse stellar streaks.

\section{GALEV modeling}
\label{galev}

To model the stellar populations in the observed star clusters and stellar streaks, we used the GALEV models of \cite{kotulla2009}, and a \cite{salpeter1955} IMF with stellar masses ranging from 0.1 to 100 $\Msun$. Recent work has suggested that the IMF in massive elliptical galaxies is `heavy' compared with less massive systems \citep{vandokkum2010, cappellari2012}, however, the form of the IMF in these regions under study is not known. Using instead a Kroupa IMF would decrease our mass estimates by a factor of $\sim$1.4.

\begin{figure*}
\centering
 \includegraphics[width=0.47\textwidth]{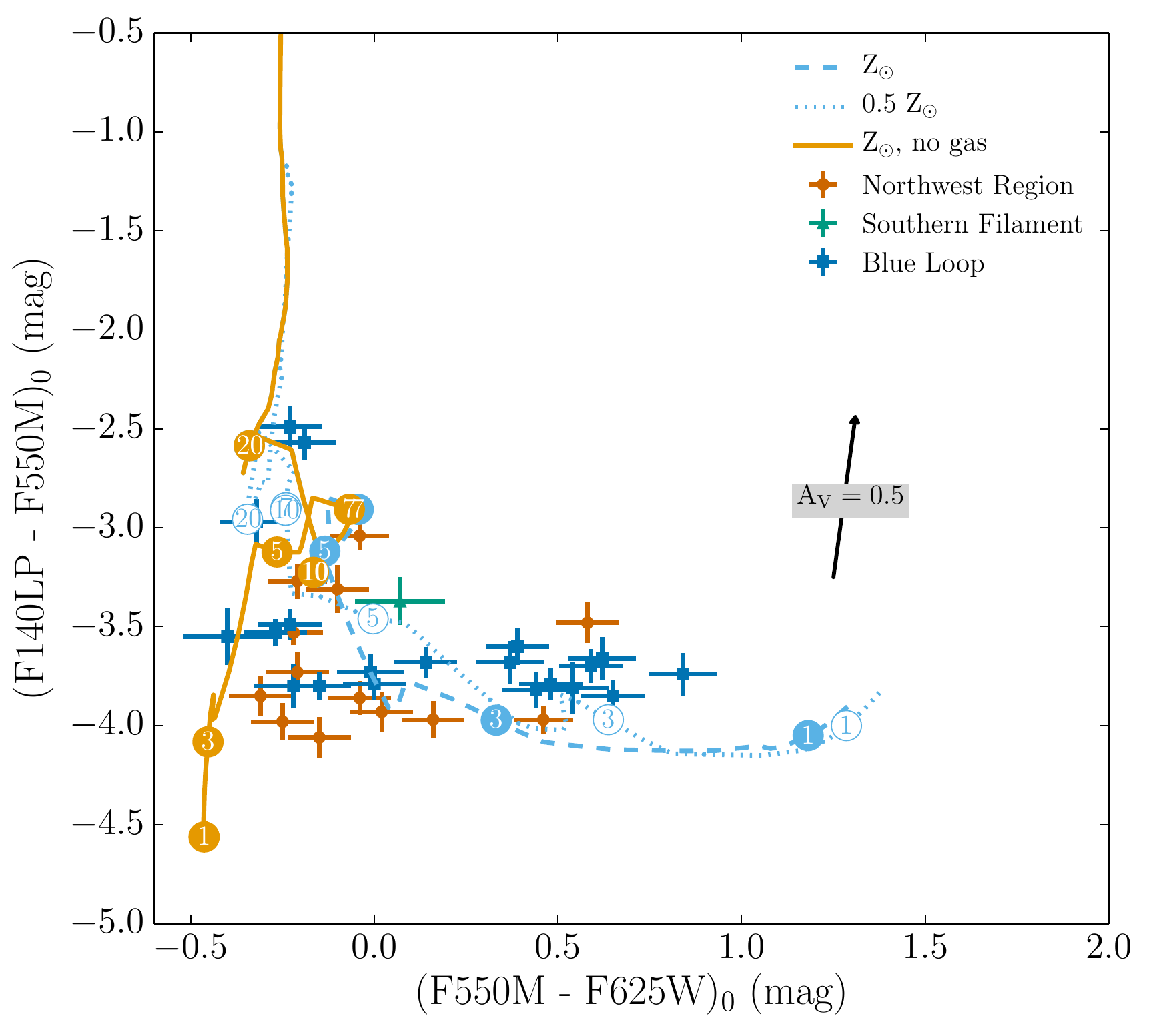}%{color-color_nogas-1.pdf}
 \includegraphics[width=0.51\textwidth]{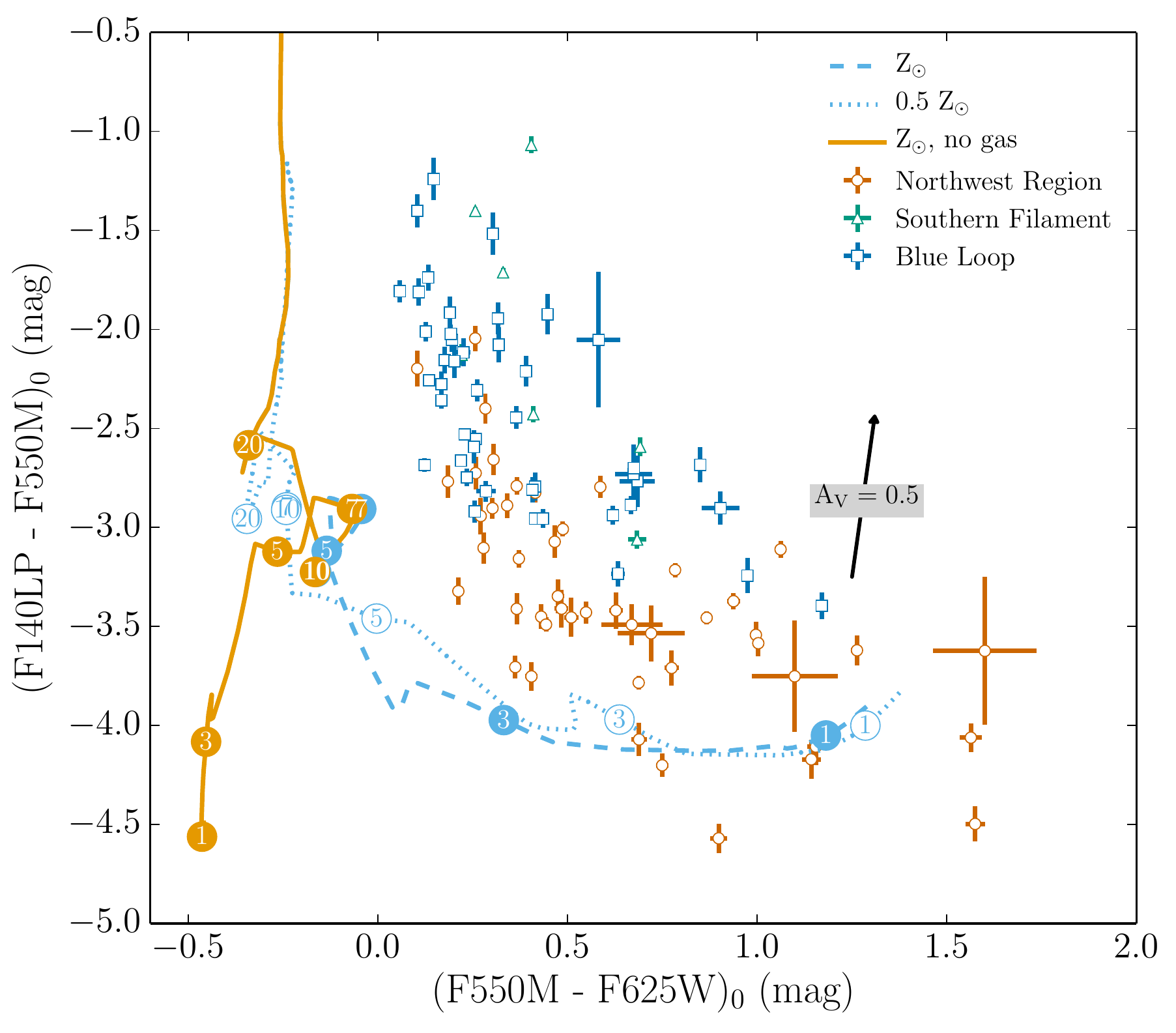}%{diffuse_colorcolor-1.pdf}
 \caption{Color-color plots of the star forming sites with foreground reddening corrections applied. The left panel shows the young clusters at the `cometary head' and the right panel the stellar streaks forming the `cometary tail' of emission. Points are the same as in Figure~\ref{cluster_cmd}. Blue dashed and dotted lines are evolutionary tracks corresponding to solar and half solar metallicity SEDs respectively, with mass $2\times10^4$ $\Msun$ (lowering the mass of the cluster has no effect on the color) and the orange solid line indicates the tracks for a solar metallicity model without line emission. The numbered points on each track represent the location of an SSP of age 1, 3, 5, 7, 10, and 20~\Myr.  \label{cluster_colorcolor}}
\end{figure*}

To obtain a sufficiently high age resolution for the expected young ages, we used the stellar evolution data presented in \cite{meynet1994}, and chose a time resolution of $10^5$ years. Our model covers the metallicity range from $1/50$ solar to twice the solar value in 5 bins, though it should be noted that our age and mass estimates are not very sensitive to the metallicity, and all models contain both gaseous and line emission. We include on some figures the models without gaseous and line emission, however, models without line emission are a poor fit to the cluster UV- or optical-F625W colors without appealing to un-physically large extinctions (see Figures \ref{cluster_cmd} and \ref{cluster_colorcolor}). Additionally, whilst most stellar streaks are offset from the H$\alpha$ filaments the UV bright `heads' often coincide in projection with the H$\alpha$ emission in the narrow-band images. These spectra for each timestep were then redshifted to $z=0.0176$ (the redshift of NGC 1275) and convolved with the filter functions for our filters, F140LP, F435W, F550M, and F625W, including the filter transmissions and detector efficiencies, to avoid uncertainties introduced by filter transformations. 

To obtain ages, metallicities, and stellar masses for each cluster, we corrected the observed multi-wavelength photometry for galactic foreground emission ($E(B-V)=0.1627\pm0.0014$ in the direction of NGC~1275, determined using the reddening maps of \citealt{schlegel1998}; corresponds to reddening values in Table~\ref{aperture_phot}) and compared it to the above model grid using the SED fitting code GAZELLE (Kotulla, in prep). GAZELLE is based on a $\chi^2$ algorithm and determines both the best-fit values and their respective 1$\sigma$ uncertainties, allowing the stellar mass and intrinsic dust extinction as free parameters. The young compact clusters are best fit with low extinctions with E($B-V$)
$\sim$0.04 and ranging between 0.0 and 0.1. Dust has never been directly detected in the outer filaments of NGC 1275. Case B line ratios indicate a relatively high H$\alpha$/H$\beta$ line ratio $\sim$3-5 \citep{hatch2005}, however, the ionization mechanism is uncertain and probably has a significant contribution from collisional excitation. When fitting the stellar streaks we make the assumption that the extinction will be similar or less than the UV bright regions and limit the extinction to be E($B-V$)$<$0.1.

\section{Properties of the Young Star Clusters}
\label{prop_star_clusters}

Paper I found a bimodal distribution of clusters in the outer envelope of NGC 1275. The bluer clusters were distributed an-isotropically, similar to, but typically offset by 0.6-1~kpc, from the H$\alpha$ filaments. In contrast the redder cluster population were distributed isotropically across the fields studied and had no morphological connection with the emission line filaments. Paper I concludes the blue population must have a different formation mechanism to the red population and the blue and red stars cannot be described as continuous star formation from a single event, the upper limits on the ages of the blue population from SSP models of the optical emission was $<$100 ~Myr. In paper I clusters were selected based on their optical properties. In this paper young compact clusters and the clumpy stellar streaks are selected by their UV properties.

Figure~\ref{cluster_cmd} shows color-magnitude diagrams (CMD) of the compact star clusters described in Section~\ref{observations}. All magnitudes and colors have been have been corrected for foreground reddening, as indicated by the subscript zeros. Evolutionary SSP tracks for a cluster with mass $2\times10^4~\Msun$, with solar and half solar metallicity and with and without emission lines, are overplotted. The numbered points represent the location of the model SSPs at ages of 1 Myr, 3 Myr, 5 Myr, 7 Myr, 10 Myr, and 20 Myr. The majority of clusters have a very blue (F140LP-F550M)$_0$ color, and lie in a narrow range of values. This plot suggests that these clusters are very young and have similar ages despite being located in physically separated regions around NGC~1275. A wider range of ages is inferred from the  optical-only CMD. Note that the tracks in the FUV-optical CMD show a simple stellar population (SSP) evolving diagonally across the plot, whereas the tracks in the optical-only CMD go back and forth over the same color range, this is due to the contributions of the line emission and red and blue super-giant (RSG/BSG) populations to the F435W and F625W filters; (F140LP-F550M)$_0$ is much more effective at discriminating ages of young star clusters than (F435W-F625W)$_0$. 

\begin{figure*}
  \includegraphics[width=0.495\textwidth]{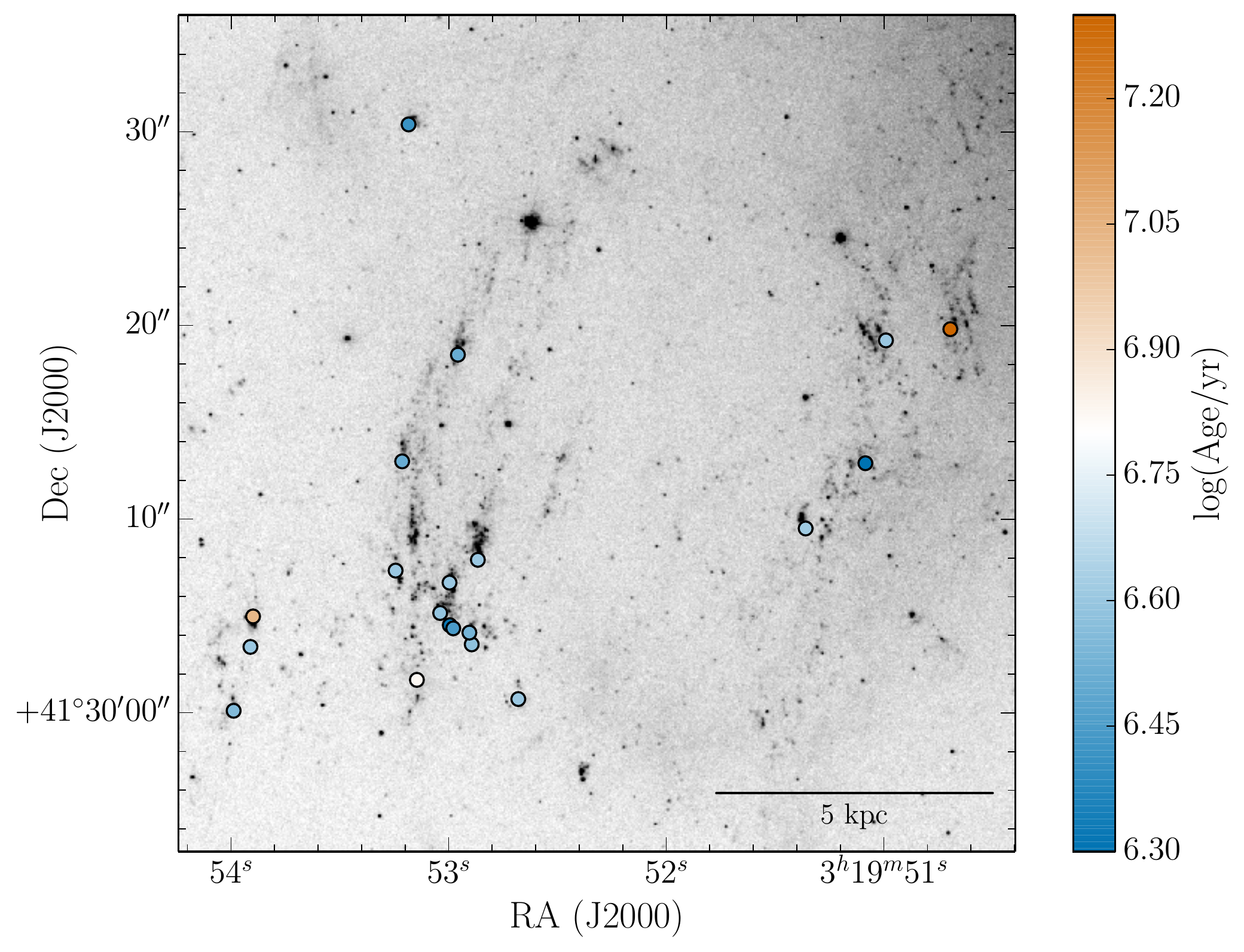}
  \includegraphics[width=0.495\textwidth]{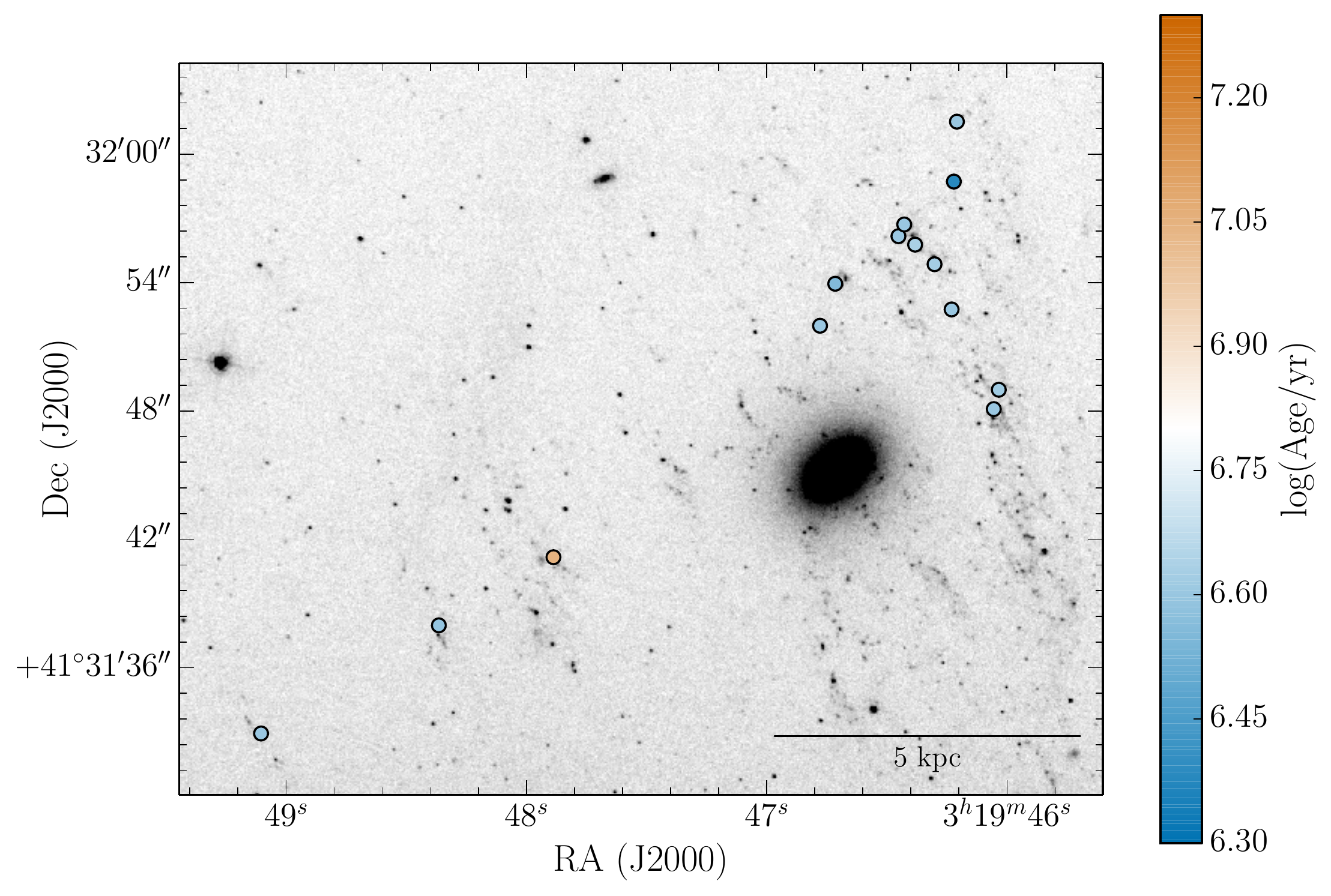}
 \caption{Ages of star clusters overplotted on a section of the F435W image which contains the Blue Loop (left) and the Northern Region (right).  \label{nr_maps}}
\end{figure*}

Figure~\ref{cluster_colorcolor} shows the (F140LP-F550M)$_0$ versus (F550M-F625W)$_0$ colors. The colors are foreground reddening-corrected, and the points and tracks are colored the same as those in Figure~\ref{cluster_cmd}. The very red (F550M - F625W)$_0$ colors of those clusters lying near the 3~Myr age indicate the presence of strong emission lines, extinction alone cannot account for the positions of these clusters. Emission lines are present in the F625W and F435W filters, whereas the F550M filter is dominated by stellar continuum. If line emission were absent, the evolutionary tracks would not extend off to the right, to the redder (F550M - F625W)$_0$ colors. Instead, the curves would be roughly vertical, beginning at (F550M - F625W)$_0 \lesssim 0$. Some clusters are present just off this location, close to the no-gas track in Figure~\ref{cluster_colorcolor}. Either these clusters have no emission-line gas associated with them but a significant extinction, or they have line emission at a lower level. Figure \ref{Ha_contours} supports this scenario; some of the compact young star clusters appear associated with H$\alpha$ emission and some are not, in every region. A fraction of the young clusters even appear to coincide with symmetrical overdensities in the H$\alpha$ emission, suggesting that the clusters may be producing HII regions. This would lend credence to our derived ages of $<$10 Myr for many of the clusters. However, we do not claim that the larger-scale emission-line filaments are produced via photoionization. Nonetheless, we do suggest that the young stellar component in the outer regions of NGC~1275 is physically connected to the emission-line filaments.
 
Figure \ref{nr_maps} shows the locations of our star clusters, color-coded for age, plotted on F435W cutouts of the Blue Loop and Northwest Region, respectively. We do not show the Southern Filament because only one compact source was identified in that region. In the Blue Loop, compact sources have an age range of $\sim$2 to 20 Myr, with an average age of 4.8 Myr, and a mass range of $\sim5\times10^3$ to $6\times10^4 \Msun$, with an average mass of $2.2\times10^4 \Msun$. In the Northwest Region, the compact sources have an age range of $\sim$2 to 11 Myr, with an average of 4.4 Myr, and a mass range of $\sim6\times10^3$ to $4\times10^4 \Msun$, with an average mass of $1.6\times10^4 \Msun$. The Southern Filament cluster has an age of 4.7 Myr and a mass of $1.8\times10^4 \Msun$. 

\begin{figure*}
 \includegraphics[width=\textwidth]{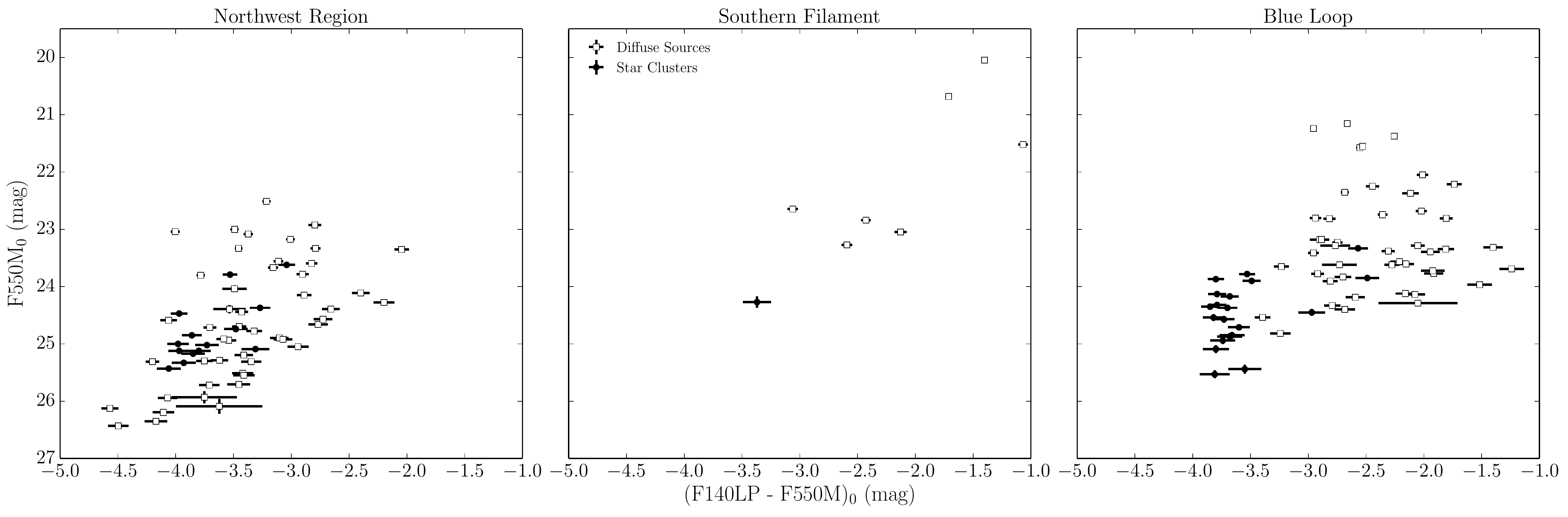}
 \caption{FUV-optical CMDs of star clusters (solid circles) and diffuse sources (open squares) in the Northwest Region (left), Southern Filament (center), and Blue Loop (right). The redder colors of the diffuse regions as compared to the star clusters imply older ages that can be estimated via comparisons to the SSP models in Figure~3. The F550M filter excludes strong emission lines, and the stellar streaks are offset from the H$\alpha$ filaments, so the redder diffuse color is not contaminated by line emission. }\label{clusterdiffuse_cmd}
\end{figure*}

\section{Properties of Stellar Streaks}
\label{prop_diffuse_sources}

As noted in Section~\ref{observations}, diffuse UV emission accompanies the star clusters in all three regions. The regions contain both very diffuse emission, presumably from star clusters which have been dispersed, and more concentrated stellar streaks with compact star clusters at their `heads' (see section \ref{prop_star_clusters}) and more diffuse emission in the `tails'.

As mentioned in paper I and can be seen in Figure \ref{Ha_contours} the `tails' of the UV streaks do not typically coincide with the H$\alpha$ emission-line filaments which surround NGC 1275. They share a similar morphology but are typically offset from the ionized gas and only appear to occur in these three disparate regions; the Blue Loop, Northwest Region and Southern Filament, whilst the H$\alpha$ filament system extends in all projected directions from NGC 1275.  The filaments have typical velocities of order 50-100~\kmps \citep{hatch2005, lim2012}, and the rotational velocity of NGC 1275 is of order 400\kmps. For the stars to remain cospatial with the filaments in our observations (a few pixels with a pixel scale of SBC $\approx$0.03''$\sim$10pc) we infer the age must be less than $\sim$1$\times10^{6}$~yrs. Therefore we may expect to see H$\alpha$ associated with the youngest clusters, as is observed, but not with the older clusters which we expect to be decoupled from the filaments. This suggests the `tails' may consist of more evolved stellar populations than the `heads'.

The streaks are orientated, in all regions, with the compact star cluster at the `head', at the farthest projected distance from the galaxy center, and the `tails' orientated towards the galaxy center, indicating balistic motion either away from or falling back towards the galaxy center. The color of the streak tails is significantly redder (see Figure \ref{clusterdiffuse_cmd}, by an average of 0.76 magnitudes) and has a larger scatter than that of the clusters. The F550M filter does not include any strong emission lines, and as mentioned above they are typically offset from the H$\alpha$ filaments, so the redder colors do not arise from emission line contamination. The tails are by definition larger in size than the clusters, which accounts for the enhanced brightness of some of the diffuse sources. Only in the Northwest Region do the tails have comparable (F140LP-F550M)$_0$ colors to the star clusters, but even here they average 0.33 magnitudes redder than the clusters. These more diffuse stellar populations may have formed outside of star clusters or originated in star clusters that rapidly dissolved. The latter scenario could explain the larger scatter in color of the diffuse sources, due to the presence of less massive, and presumably older, stars \citep{tremonti2001,chandar2005}.

\begin{figure}
\centering
 \includegraphics[width=0.5\textwidth]{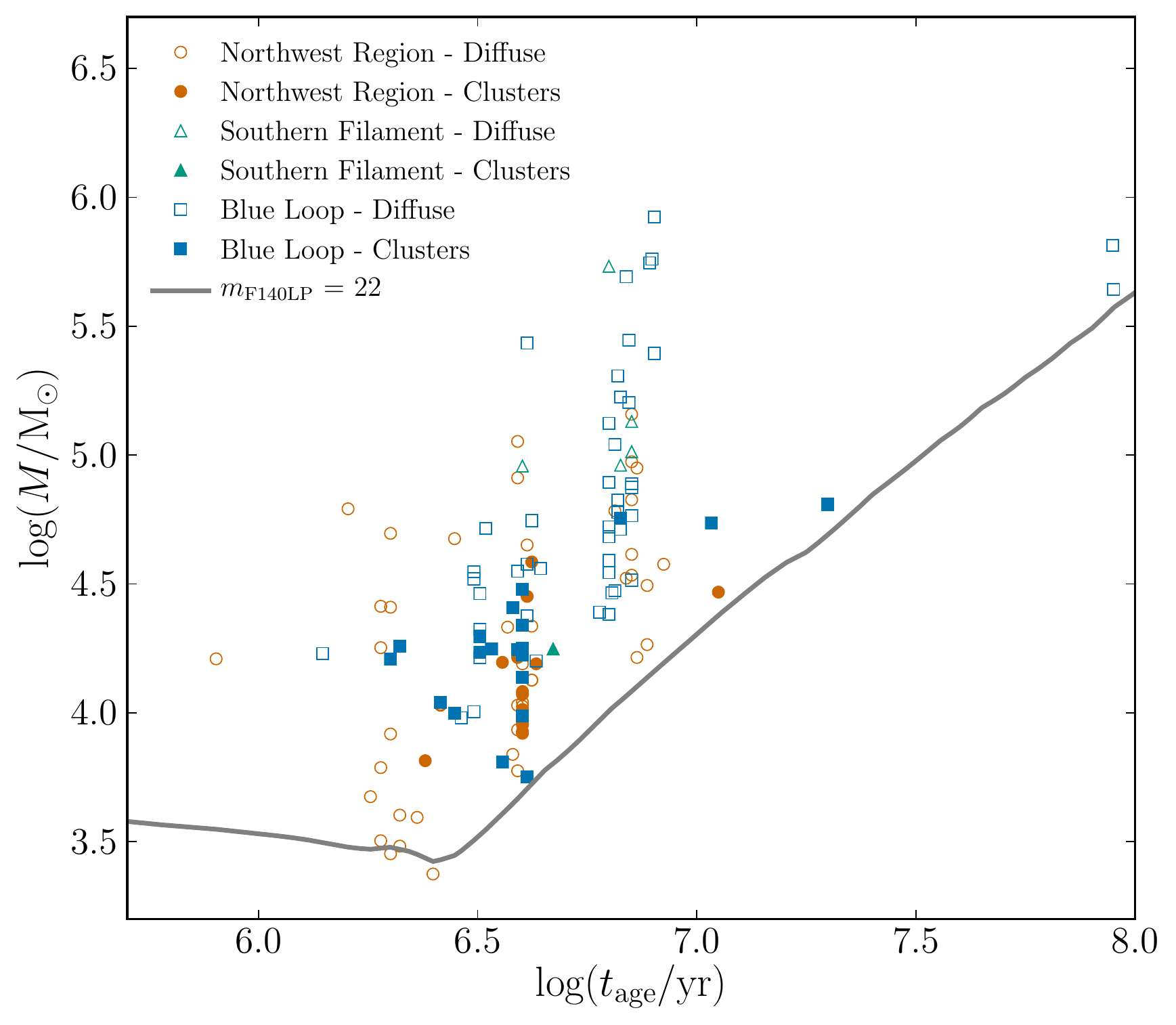}
 \caption{Masses and ages of star clusters (solid points) and diffuse sources (open points). The blue squares are sources in the Northwest Region, the green triangles are sources in the Southern Filament, and the red circles are sources in the Blue Loop. The solid line shows a detection limit of F140LP $= 22.0$ magnitudes, which shows how star clusters fade as they age. The clumping of ages is an artifact of the fitting method, see section \ref{galev}. \label{age_mass}}
\end{figure}

Assuming SSP models, ages of the tails range from a few to a $\sim$50 Myr, with an average age of $\sim$8 Myr, thus they tend to be older than the compact star clusters (see Figure~\ref{age_mass}). They have stellar masses in the range of a few 10$^3$ to 10$^6$ M$_{\odot}$ with a total mass in all regions in the range 1$\times$10$^{7}$-8.6$\times$10$^{7}$~\Msun. Very recent star formation, at a rate of a few to$\sim$13~\Msunpyr, has taken place over the three distinct regions.
These estimates are derived under the assumption that the diffuse regions do not systematically experience more interstellar obscuration due to dust than the star clusters, however, we note here that the best fit average extinction of the diffuse regions is twice that of the compact clusters ($E(B-V)\sim0.08$)
which, while not a significant effect, may be biasing our ages for the diffuse clusters marginally low. We note also that each of the `tails' are unlikely to be a SSP and probably consist of stars with a range of ages, with a significant contribution to the blue light from the youngest stars. This is highlighted by the color-color plot in Figure \ref{cluster_colorcolor} which shows that for the `tails' the data points are redder than the SSP models allow. The redder colors may also be due to a much greater level of extinction in the tails of the streaks compared with the heads but this seems unlikely due to the large radius from the center of the galaxy and offset from the gaseous filaments. The stellar populations making up the `tails' are likely formed from stellar populations of stars ranging from a few to $\sim$50 Myrs. The total mass in the stellar streaks is of the order $\sim$5$\times$10$^{6}$\Msun\ in the Blue Loop and Southern Filament, and $\sim$10$^{6}$\Msun\ in the Northwest Region, a factor of 10 larger than the mass in the compact clusters.
A very diffuse component to the FUV emission also exists so the mass of young stars estimated in these regions is a lower limit. This is consistent with a picture whereby the tails of the streaks are made up of older stars as proposed in Paper I. However, large errors in the age estimations, due to the lack of a UV filter, meant the authors were unable to detect an age gradient.

\begin{figure}
 \includegraphics[width=0.5\textwidth]{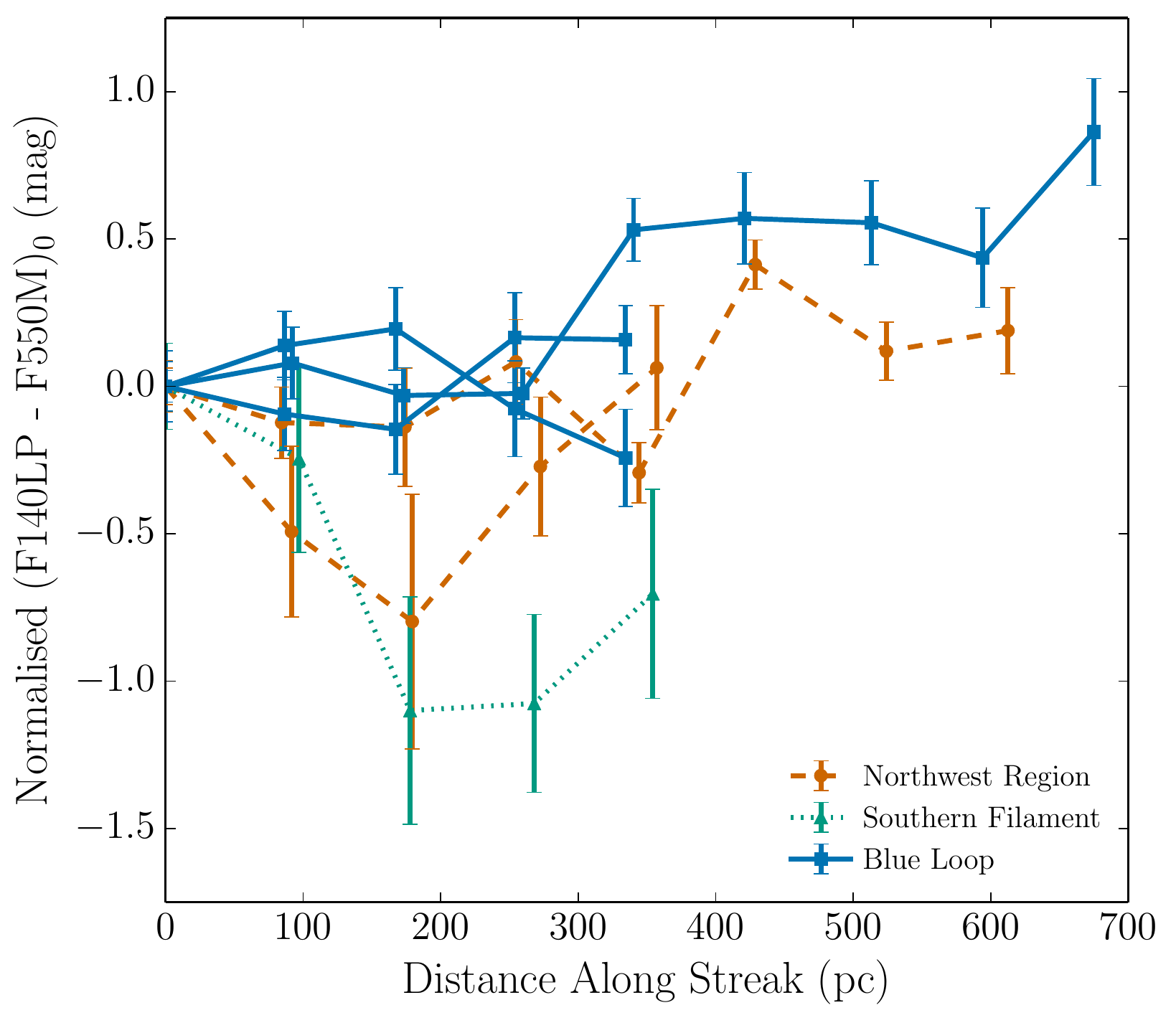}
 \caption{Normalised HST color (F140LP - F550M) gradient along five streaks. The blue lines are for three streaks in the Blue Loop, the green line is for one streak in the Southern Filament, and the red lines are for two streaks in the Northern Region.} \label{age_grad2}
\end{figure}

To investigate whether an age gradient exists along the stellar streaks, we measure the fluxes in the four HST bands in apertures of radius 0.2'' along selected streaks in all three regions. Many streaks are inherently clumpy, complicating the definition of the ``head'' and ``tail'' of the streak. The best examples of smooth stellar streaks are observed in the Blue Loop region. No aperture corrections are performed. We are interested in whether there is a relationship between distance along the streak and the color of emission, not absolute magnitudes. In addition, our apertures are much larger than the PSF and the star clusters, so aperture corrections should not affect our conclusions. Figure~\ref{age_grad2} shows the results of this photometry with the `heads' of the streaks at a distance of zero, note however the apertures used here are larger than those used in the stellar cluster analysis. Though an age gradient may exist in the Blue Loop and Southern Filament, the errors in photometry are still too large to confirm a gradient for many of the streaks. 
However, within the same streaks the heads are approximately one mag bluer than the tails across all regions. These color differentials suggest the tails are not simply debris from the clusters at the heads of the streaks.

\section{Discussion}
\label{discussion}

The H$\alpha$ filaments, some as long as $\sim$70~kpc, are distinctive features of the LVS of NGC~1275, the central galaxy in the Perseus cluster.  The filaments are multi-phase structures, readily detected via line emission from the $\sim$10$^{4-6}$~\Msun\ of ionized gas \citep[e.g.,][]{conselice2001} but also concealing $\sim5\times$10$^{10}$~\Msun\ of molecular gas within 50~kpc of the nucleus \citep{salome2011}. Despite this large cold gas reservoir, only three disparate regions of the filamentary system support high levels of star formation, while the majority of filaments remain relatively quiescent \citep{conselice2001, hatch2006, johnstone2012}. 

The morphologies and exceedingly young ages of the star forming regions studied here, suggest that the stellar streaks must be formed in situ from the filaments.  Our FUV analysis further shows that when star formation does occur, it takes place rapidly. We find well organized features delineated by stellar populations with ages of $\leq$50~Myr and with typical ages of only a few Myr in the bright compact clusters. The three regions studied have remarkably similar blue colors with little scatter so are all currently forming stars. Paper I, using optically detected clusters, showed that the star formation in these filaments had an age $<$100 Myrs but the optical colors were unable to constrain the ages better than this. These optically detected clusters lie in the `streaks' discussed here with ages of less than or a few 10's of Myrs.  A model for the outer star formation in NGC~1275 will need to explain the dichotomy between the quiescent and star forming filaments, the disparate nature of the star formation, as well as the details of the star formation process.  In this section we present first steps towards addressing these issues.

\subsection{Quiescent Filaments}
\label{quescents}

The majority of the filaments in the NGC 1275 LVS, are quiescent.  The filaments are likely to be products of disturbances produced in a high pressure, multi-phase interstellar medium by buoyant bubbles of relativistic plasma produced by the central AGN \citep[e.g.,][]{fabian2003a}.  Lifetimes for filaments are estimated to be $\approx$10$^8$~Myr based on crossing times derived from their $\sim$200~km~s$^{-1}$ peculiar velocities with respect to the center of NGC~1275 and on bubble rise times derived from observations of holes in the X-ray gas \citep{dunn2005}.

The existence of dense gas filaments is puzzling given the location of NGC~1275 at the center of the Perseus galaxy cluster, with its high pressure, hot intracluster medium \citep[e.g.,][]{fabian2000}.  Filament survival for at least a dynamical time scale is not understood, but requires that the filament gas can cool with sufficient efficiency as to prevent rapid evaporation.  Thus \cite{fabian2008} emphasize the importance of magnetic fields in defining and also insulating the filaments from excessive conductive heating from the surrounding hot gas.  Some heating by the hot gas, however, does take place \citep[e.g.][]{ferland2009}, and may also contribute to the large measured turbulent velocities, e.g., $\sigma \approx$50-100~km~s$^{-1}$ that are observed within the inner filaments \citep{hatch2006, salome2008, salome2008a}. It should be noted, however, that the turbulent velocities must include the effects of observing multiple filament threads in projection.

We therefore hypothesize that the existence of filaments in NGC~1275 requires special conditions.  An increase in power absorbed per unit mass over time or loss of shielding magnetic fields in the filaments could result in evaporation and disruption.  On the other hand if cooling triumphs, star formation could result.  

While young stellar populations exist within the central $\sim$15~kpc, in NGC 1275, these clusters are significantly older than the populations detected in the filaments at radii $>$15~kpc, and their spatial distribution does not correlate with that of the filaments \citep{carlson1998, canning2010a, penny2012a}.  Observations of the distribution of young stars relative to the ionized filaments suggests a simple model: Heating dominates for the inner filament system, preventing star formation, and it is only in the outer regions of NGC~1275 that some filaments can become unstable against large scale star formation. 

In suggesting this empirical model, we are putting aside the issue of the longer term evolution of non-star forming filaments.  Do these systems evaporate, or fall back into the galaxy to be mixed into new filaments by turbulent processes?  Since star formation occurs in only a minority of filaments, some process in addition to star formation must limit the radial extent of the LVS filaments.

\begin{figure}
  \centering
  \includegraphics[width=0.5\textwidth]{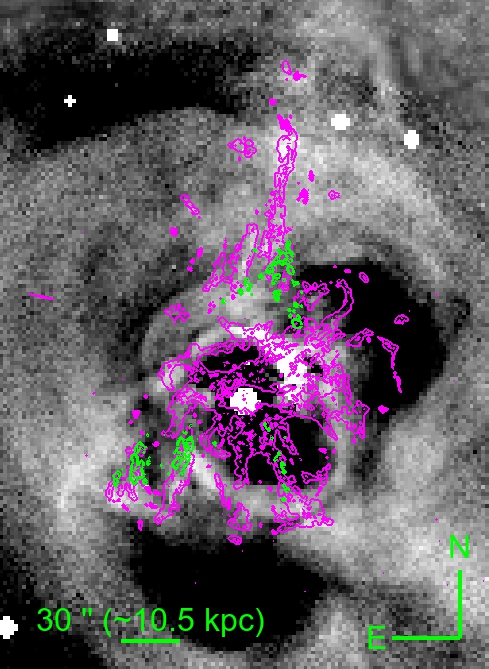}
  \caption{{\it Chandra} X-ray surface brightness image from \citep{fabian2011a} with the average at each radius subtracted overlaid with contours of H$\alpha$ emission (magenta; \citep{conselice2001}) and young star forming regions (green; this study). The X-ray bubbles blown by the AGN are seen as dark depressions in the X-ray surface brightness image. The center of NGC 1275 is at coordinates 49.9507$^{\circ}$, 41.5118$^{\circ}$.}\label{xray_ha_uv}
\end{figure}

\begin{figure*}
  \centering
  \includegraphics[width=0.4\textwidth]{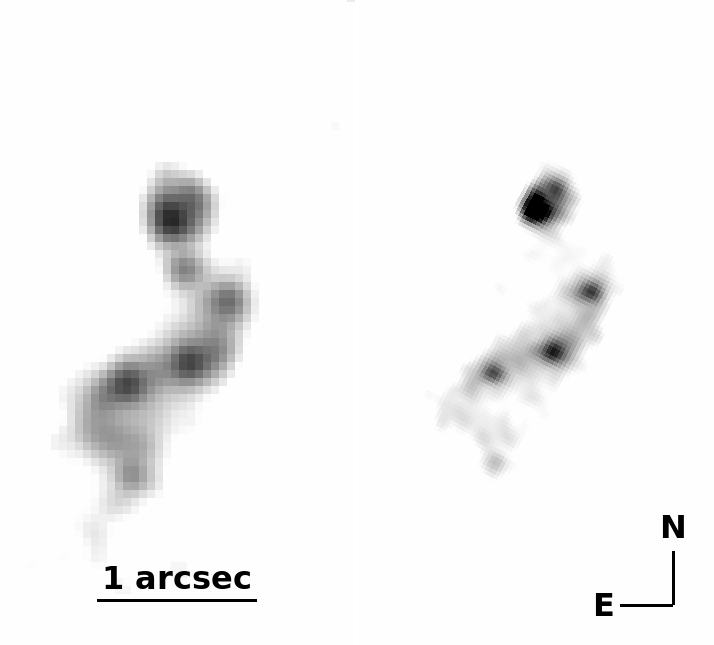}
  \includegraphics[width=0.4\textwidth]{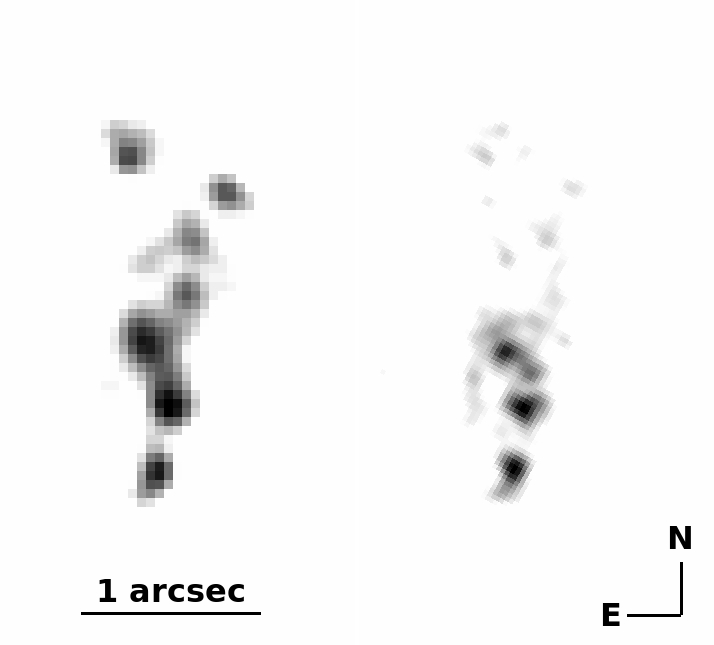}
  \caption{The ``Snake' (49.9408$^{\circ}$, 41.5527$^{\circ}$)' {\bf (left)} and a section of the Blue Loop (49.9706$^{\circ}$, 41.5027$^{\circ}$) {\bf (right)} in F435W and F140LP emission. Star formation in stellar streaks occurs in clumps with a projected periodicity of $\sim$0.3-0.5 arcsec (105-175pc). }\label{periodicity}
\end{figure*}

\subsection{Filaments and Star Formation}
\label{gas_filaments}

\subsubsection{External Perturbations}

The ICM is not static; bulk sloshing motions which spiral to large radii can be seen as spirals in the intracluster medium X-ray surface brightness, in the spectacular, deep {\it Chandra} X-ray images presented in \cite{fabian2011a}.  The sloshing motions are likely due to a subcluster merger or interaction and can persist in the ICM for a very long time \citep[for a review, see][]{markevitch2007}.  However, the bulk motions do not appear to be effective triggers of star formation in the filaments as shown by the radial extent of the filaments, smooth radial velocity measurements \citep{hatch2006} and long lifetimes.

Rising relativistic gas bubbles might also affect pre-existing filaments, possibly leading to star formation. Figure~\ref{xray_ha_uv} shows the projected spatial correspondence between the X-ray surface brightness, ionised gas filaments, and UV emission in the three regions under study. The correlation between the ionised gas filaments and the star formation is obvious, but any correlation of these features with the X-ray bubbles is less clear: In the southeast, the centre of the Blue Loop coincides spatially with a depression in the X-ray image. The expansion of this bubble underneath the ionised gas could be triggering compression and star formation along a filament. However, no clear correspondence of SF with the X-ray bubbles is observed in the Northwest Region or Southern Filament.  The star formation in the Southern Filament is slightly offset from nearby ionised gas filaments, and is confused with the large inner bubbles which dominate the X-ray signatures in this region, while in the Northwest the star formation is not clearly associated with any bright dense regions of X-ray gas, although X-ray depressions are apparent to the east, west, and south.  It is also important to note that we do not know the projection angles of the ionised filaments or of the star-forming regions.  Additionally, little scatter exists in the ages of the compact stellar features across all regions, with star formation switching on $<$50 Myrs ago, suggesting a common mechanism disrupted the three filaments. Estimated rise times for the large southern-most bubble preceding the blue loop, and the large bubbles still connected to the nucleus (see Fig. \ref{xray_ha_uv}) differ by close to order of magnitude being 75~Myrs and 15~Myrs respectively \citep{dunn2006}. If separate generations of X-ray bubbles were responsible for the star formation this would lead to detectable variations in the ages of the clusters in each region.  However, estimations of bubble rise times can have large uncertainties from projection effects. Additionally these bubbles use different methods to estimate their age as the inner bubbles are still connected to the nucleus. The buoyancy timescale is used for the outer bubbles and the sound speed timescale (an expansion timescale) for the inner bubbles (see \citealt{dunn2006}).

Weak shocks, from the generation of the bubbles, are apparent in the X-ray gas and could simultaneously disrupt the filaments.  The straight edge running northwest through the Northwest Region may be caused by a weak shock in the gas from the inflation of the bubbles. The projected distance from the nucleus of $\sim$25~kpc would require a transport process travelling at $\sim$1000\kmps to synchronise the star formation to tens of Myrs.  A shock in the hot X-ray gas might reach these velocities.  However, a shock would compress the gas along the shock front, not perpendicular to it and should have disrupted also the inner regions of the filaments leading to a gradient in age with distance from the nucleus, we don't find any evidence for this in our data.

While the structure of NGC~1275 resembles that of late phase field E-galaxy merger products with shell-like features (e.g., \citealt{schweizer1998}), these features are unlikely to originate from simple gas-rich mergers. Wet mergers with the BCG are unlikely due to the high velocity dispersions in massive, low redshift galaxy clusters; the active star formation in NGC~1275 probably does not result from the kinds of processes occurring in wet mergers of field galaxies (\citealt{monaco2006, conroy2007}). An external perturbation such as disruption of pre-existing filaments by a passing galaxy or dark matter clump is not a likely mechanism for producing the outer young stellar filaments.  Any effects from such collisions, such as tidal compression, should be localized around the site of the interaction, which might affect regions in a few filaments, but not yield large coherent features.  Tidal debris from an infalling star-forming galaxy, as in the high velocity emission line system (HVS), would likely produce more spatially localized and disturbed young stellar structures than what we observe.  Furthermore, the galaxy responsible for the tidal debris should be observable.  Aside from the obvious presence of a galaxy in the HVS, the debris from which is unrelated to the regions we are studying, no candidate impactor has been found that could simultaneously lead to star formation in the southeast, south and northwest of NGC~1275.

\subsubsection{Internal Instabilities}
\label{instabilities}

As discussed in Paper~I, the structure of the young stellar streaks is suggestive of coherent motions. We also note that the streaks within a given region tend to be parallel to one another, as expected if the stellar streaks are moving along similar paths. With our more accurate age estimates, we can estimate the minimum velocity of the streaks if they are associated with progressive star formation. Projected streak lengths are $\sim$0.7-1.5~kpc, which gives $v_{proj} \approx 30/t_*(20)$ to $60/t_*(20)$~km~s$^{-1}$, where $t_*(20)$ is the stellar population age range in units of 20 Myr. 

This suggests that the speeds of the young stars could be slower than the 100-200~\kmps radial velocities measured  for outer filaments by \cite{hatch2006}. Paper~1 discusses the possibility that once stars form in the filaments, they can no longer be supported by gas and/or magnetic pressure. If star-forming filaments stall in the outer parts of NGC~1275, stars and star clusters could begin to fall back towards NGC~1275, forming the stellar streaks we observe. Another possibility is that the streaks represent the outward velocities of the gas filaments, but in either case the parallel streaks within the regular large scale patterns of star formation indicate that the filaments responsible for star formation retained much of their initial structure during the star formation process, implying that star formation was not triggered by a disruptive process.

We first consider the possibility that the star formation is due to perturbations propagating along filaments. Our GALEV models show that the spread in star cluster ages is $\ll$50~\Myr. Assuming a filament length of 10~\kpc, any triggering perturbation would need to propagate along the filament at speeds of $\geq$200~\kmps. The sound speed for gas with 10$^{4}$~K, typical of the ionized component of the filaments, is $\approx$15~\kmps, while in the cold dense molecular gas (T$\sim$100~K) the sound speed is $\sim$2~\kmps. Any disturbance propagating along a filament with sufficient speed to produce the observed small range in stellar ages would need to be highly supersonic and would lead to observable shock line diagnostics which are not seen. 

\cite{ferland2009} and \cite{fabian2011b} suggest that if the excitation of the filaments in NGC~1275 is from energetic particles in the hot gas impacting and mixing with the cool and cold gas phases, these particles may lose energy and remain in the filament. In this way the filaments, which were initially seeded through gas uplift, may grow in mass by up to 100~\Msunpyr.  Filaments may then grow in mass to the point where they become Jeans unstable and fragmentation of the filament leads to massive star formation.  However, near simultaneous growth of only three disparate filaments is again difficult to place within this scenario, we might instead expect to observe continuous outer halo star formation across all the filaments.

A very rough periodicity exists in the spacing of the streaks along a filament, with typical spacings of $\sim$1~kpc.  This pattern resembles the ``beads-on-a-string" distribution of star formation along spiral arms, which result from large-scale gravitational instabilities \citep{elmegreen1983, kim2006}.  On smaller scales, some stellar streaks contain several star clusters, also spaced at somewhat regular intervals of roughly 100~pc ($\sim$0.3~arcsec), though projection effects make it difficult to measure the exact wavelength of perturbations.  Figure~\ref{periodicity} illustrates this periodicity with close up  F435W and F140LP images of two such streaks, the ``Snake'', as mentioned in Section~\ref{observations}, and part of the eastern side of the Blue Loop. Similar periodic structures are also seen in nearby star forming filaments such as the `Nessie Nebulae' \citep{jackson2010}.  Modelling the filaments as infinite, isothermal cylinders, supported by thermal pressure in the gas, with a central density 10$^{4}$\pcmcu, the periodicity of star formation is expected to be $<$10~pc \citep{chandrasekhar1953, nagasawa1987}, which would not be observable in our data. Magnetic fields and turbulence in the filaments can add support and increase the maximally unstable wavelength. If turbulent support dominates then turbulent motions of $\sim$17~\kmps are required for the observed cluster spacing. This is comparable to the limit of current observations which are limited by the beam size, however, high spatial resolution observations may allow us to quantify the turbulent pressure support in these structures. NGC 1275 is at the limit of observability by ALMA. These structural features indicate that gravitational instabilities within filaments are playing a key role in the outer star formation in NGC~1275, although the dynamics of substructure formation likely differs between spiral arms and filaments \citep[e.g.,][]{shetty2006}.

\subsection{Star Formation in Unstable Filaments}
\label{model}

We therefore consider models where stars form in response to the overall evolution of some gas filaments. Our conceptual model assumes that the ionized gas filaments maintain their identities as they expand outward from the center of NGC~1275. This assumption is consistent with the structures of ionized filaments observed in high angular resolution {\it Hubble Space Telescope} images, with dynamical models for filament formation associated with rising bubbles of relativistic material ejected from the AGN \citep{fabian2003a,hatch2006,fabian2008}, as well as with models invoking large scale turbulence within NGC~1275 to organize matter into filaments \citep{falceta-goncalves2010, fabian2011b}.

Ionized gas within filaments located in the main body of NGC~1275 frequently has large internal velocity dispersions ($\sigma \approx50-100$~\kmps), while in the outer galaxy dispersions appear to drop to $\sigma \sim$20~\kmps \citep{hatch2006}. Whilst confusion with multiple filament threads may play a significant role here, one explanation for this behavior is that gas filaments in the inner galaxy are stabilized against star formation by their high internal velocity dispersions perhaps maintained by the heating and turbulence generated by the radio mode AGN. Those filaments that reach the outer galaxy with low internal velocity dispersions and sufficient molecular masses may be subject to gravitational collapse along the filament, which leads to star formation. Several conditions must be satisfied for this to occur, including a sufficient mass density, an ability for the molecular gas to cool to temperatures typical of star-forming regions ($<$100~K), and that magnetic pressure does not prevent collapse. Zhuravleva et al. in prep. explore the X-ray surface brightness fluctuations in the core of the Perseus cluster and find stronger fluctuations towards the center possibly indicating higher turbulence at smaller radii (see also \citealt{zhuravleva2014, gaspari2014}). The forthcoming Astro-H satellite will be instrumental in constraining turbulent motions in galaxy clusters.

The collect-and-collapse mechanism for star formation in expanding shells \citep{elmegreen1977} has some features in common with such a filament. As an expanding shell sweeps up gas from the surrounding medium, it can become gravitationally unstable against star formation \citep{dale2009}. These instabilities cause the shell to break up into clumps of gas, but the end result  depends on the relative mass in these structures.  If there is sufficient mass density in the perturbed regions, near-simultaneous star formation can occur in the shell over short timescales. This phenomenon is seen, for example, in star formation associated with supershells in the Large Magellanic Cloud \citep[e.g.,][]{efremov1998}, and in NGC~2146, where a central star formation event has driven a shell into the surrounding dense molecular medium, causing the formation of a ring of star clusters \citep{adamo2012}. 

In NGC~1275, however, the filaments expand into the hot, low density ICM, which prevents them from sweeping up a substantial gas mass. The standard shell instability probably does not apply.  Fortunately, the case of a thin gaseous shell expanding into a high pressure, low density medium has also been studied \citep{dale2009}. Due to the presence of the hot ICM, the pressure within a $\sim$75~kpc radius around NGC~1275 stays constant to within a factor of $\sim$2 \citep{fabian2007}.  We can qualitatively consider whether a pressure-induced instability might apply in an outer NGC~1275 gaseous filament where the internal velocity dispersion is low. While models refer to spherical shells dominated by thermal gas pressure, the filaments around NGC~1275 experience externally-driven heating and turbulent velocities \citep[e.g.,][]{fabian2011b}, as well as substantial magnetic fields \citep{fabian2008, taylor2006b}.  A full  theoretical study therefore is needed to see if this approach quantitatively applies to the situation in NGC~1275.

A massive, thin gaseous shell expanding into a vacuum that is initially gravitationally stable may remain in this state. As the shell expands, its density declines, which inhibits the growth of gravitational instabilities. Some combination of increased gas density, reduced expansion rate, and decreased pressure timescales is required for the shell to break up into self-gravitating clumps, which are potential sites for star formation. The pressure assisted gravitational instability \citep[PAGI;][]{wunsch2010,wunsch2012,dale2009} is a possible candidate for producing self-gravitating sites for star formation from outer filaments in NGC~1275. In PAGI, the swept-up mass is assumed to be small, thereby reducing effects of the Vishniac instability so that the filament could remain in equilibrium with a substantial external pressure.  Under these conditions even a filament that is stretching as it moves outwards can sustain a constant or even decreasing thickness so as to meet the requirement of pressure balance.  Thus the filament density may not significantly decrease relative to a freely-expanding thin shell, and may even increase.  Higher gas densities in turn promote Jeans-types of gravitational fragmentation with the passage of time.

Our initial picture for outer star formation in NGC~1275 is that a gaseous filament, perhaps with a denser or less turbulent molecular medium than in the majority of filaments in the main body of NGC~1275, coasts radially outwards over a time scale of  $\sim$10$^8$~yr . As a filament leaves the inner galaxy, a combination of factors can act to increase its density. One possibility is that the modest drop in density of the surrounding hot medium leads to PAGI, which could be aided by decreases in the gas temperature and turbulent velocity within the outward moving filament. An additional factor may come from filament dynamics. Filament expansion may slow in the outer parts of NGC~1275 due to the galaxy's gravitational potential. Thus the shear that occurs if a filament is being stretched in radius, as in the \cite{fabian2003a} model for filament interactions with rising bubbles of non-thermal radio-emitting lobes, would decrease.  The magnetic field also may become less of a factor over time. Once the gravitational instability becomes non-linear, the molecular gas dominating the filament mass  breaks up into classic ``beads on a string'', producing roughly equally-spaced sites of star formation making streaks, within which the spaced out star clusters are found. Star formation within the gravitationally unstable region of a filament then occurs over time scales of tens of Myr, leading to star forming sites being elongated into streak-like structures.

\subsection{Filaments and Stellar Halo Formation}

Once stars form they will move on ballistic orbits within the gravitational potential. Since the NGC~1275 filaments show at most modest signatures for rotational support, the resulting stellar orbits will inherit low specific angular momentum and so may pass through the main body of NGC~1275. Even so they will spend most of their time in the outer parts of the galaxy.  Thus in NGC~1275 we observe the production of an outer stellar halo due to a specific mode of star formation, rather than from collisional stellar debris associated with galaxy-galaxy interactions.  Our inferred star formation rate of $\sim$2-3~M$_{\odot}$~yr$^{-1}$ is sufficient to produce a significant stellar population around the main body of NGC~1275, provided that the process will remain active for at least several hundred Myr.  Many generations of X-ray bubbles are observed in NGC 1275; if these bubbles are responsible for depositing gas reservoirs in the outskirts of the galaxy then this process has been occurring for $>$5$\times10^{8}$~yr \citep{fabian2011a}.

Is star formation and star cluster formation in gas filaments arising from the interiors of galaxies a unique feature of BCGs in rich galaxy clusters?  While similar gas filaments are found around many giant elliptical galaxies with large X-ray halos and hosting powerful radio-mode AGN, only in rich cool-core cluster BCGs have the filaments been found to extend beyond $\sim$10$-$15~kpc.  These BCGs are surrounded by high pressure atmospheres and require more powerful feedback from the AGN to balance the cooling X-ray gas.  During galaxy formation, however, a pressurized external atmosphere may be supplied by infalling cosmic gas, and powerful AGN are more common.  We therefore wonder if star formation modes related to that seen in NGC~1275 could be important in producing the stellar halos of galaxies with their systems of globular clusters?  \cite{penny2012a} find an abundance of ultra-compact dwarfs (UCDs) concentrated around NGC 1275 in the Perseus cluster, at least two of which are associated with the filaments \citep{penny2014}. UCDs may be the high mass end of GCs. They speculate that the bluest UCDs may be formed from the extensive filament system.  The early formation of galaxies seems an excellent opportunity to produce expanding shells and filaments of dense gas that reach significant radii from their protogalactic centers. Star formation in these systems could lead to dynamically hot, spatially extended stellar halos and globular cluster systems.

\section{Conclusions}
\label{conclusions}

Our {\it Hubble Space Telescope} FUV observations of three outer regions ($R_{\mathrm{projected}}\sim$20-45~kpc) of NGC~1275, the luminous central galaxy in the Perseus cluster, confirm the presence of products of recent star formation with ages of $\leq$20~Myr. We find ``streaks'' of young stars with typical size scales of  $<$1~kpc. Embedded within some of these streaks are compact sources, which we find are consistent with moderately massive star clusters. The young star clusters and stellar streaks form filament-like, large-scale structures that are spatially and morphologically correlated with, yet offset from, some of NGC~1275's famous giant gas filaments, and therefore are associated with the galaxy (i.e., they are in the low velocity system, LVS).  The 10~kpc scale Blue Loop is a prime example of this phenomenon. While the majority of the gas filaments in the main body of NGC~1275 appear to be free of significant star formation, a minority of filaments evidently reached the outer zones of NGC~1275, where they produced galactic-scale star-forming structures with a surprising degree of coevality within each large scale structure.
 
Our data supplement an initial photometric analysis of the stellar populations in the young star clusters and stellar streaks presented in Paper~1. We estimate the mass in stars formed in the observed areas during the past $\sim$10~Myr to be somewhat greater than 10$^7$~M$_{\odot}$, implying a star formation rate of $\geq$2-3~M$_{\odot}$~yr$^{-1}$. While this is a small fraction of the total star formation rate in NGC~1275, $>$10~M$_{\odot}$~yr$^{-1}$, it represents a substantial contribution to the stellar content of the galaxy's outer halo. Only  about $7 \times 10^5$~M$_{\odot}$, or less than 10\% of the young stellar mass in the regions we observed, is located in star clusters. Therefore, the cluster formation efficiency in these regions appears to be typical for low intensity star-forming regions \citep[e.g.,][]{goddard2010}. However, it is surprising that some star clusters are being formed by gas injection into isolated regions with very low mass densities.  Once stars form, any gas or magnetic pressure support are removed, suggesting that the young stars and star clusters will move ballistically in the gravitational potential of NGC~1275 and the Perseus cluster core. We suggest that this dynamical effect produces the observed streaks of star formation, although we are unable to determine if the streaks represent an initial outward motion of the gas filament, or are due to recently formed stars dropping out of a stalled gas filament and falling inwards on ballistic orbits.  

The combination of morphologies and coincident locations demonstrate that the star-forming regions studied here are connected to the H$\alpha$ filaments in NGC~1275. These are likely to be energized by the injection of over-pressured regions of relativistic particles from the AGN, which produce expanding buoyant bubbles. The bubbles interact with the interstellar medium to produce dense filaments.  The filaments in NGC~1275 frequently contain molecular gas, and therefore are primed for star formation, but usually fail to do so. We suggest that the lack of extensive star formation in the majority of filaments located within the main body of NGC~1275 could be related to high internal turbulent velocities in combination with significant magnetic pressures.  

Filaments that retain their molecular medium, expand outwards, and slow down, in the nearly constant pressure ICM may become gravitationally unstable, especially if internal velocity dispersions drop and magnetic fields cease to be dynamically  dominant. This leads to circumstances akin to the pressure-assisted gravitational instability hypothesized to operate in thin shells, and could lead to gravitational collapse and subsequent star formation. The situation in NGC~1275, however, is complex, e.g., due to the likely importance of magnetic fields in the inner filaments.  Further study is required to properly understand why some filaments are able to form stars while others do not, as well as why extensive star formation appears to be a property of filaments that reach the outer parts of NGC~1275.

This type of mechanism, star formation in expanding shells or filaments of material, might offer a method to produce dynamically hot stellar populations and star clusters in young galaxy halos, such as globular star cluster systems. Since globular clusters are thought to have formed in the first few Gyr of the lifetime of a galaxy, they would have been produced in an era when galaxies and their surroundings were gas rich and AGN activity levels were high. This environment seems a fertile ground for making stars and star clusters via mechanisms that we see operating today, albeit under very different circumstances, in NGC~1275.

\section{Acknowledgments}

We thank the referee for helpful comments. J.E. Ryon acknowledges the support of the National Space Grant College and Fellowship Program and the Wisconsin Space Grant Consortium. R.K. gratefully acknowledges ﬁnancial support from STScI theory grant HST-AR-12840.01-A. Support for Program
number HST-AR-12840.01-A was provided by NASA through
a grant from the Space Telescope Science Institute, which is
operated by the Association of Universities for Research in
Astronomy, Incorporated, under NASA contract NAS5-26555. This research was supported in part by
grant GO-11207 from the Space Telescope Science Institute.

\bibliography{mnras_template}

\end{document}